\documentclass[twocolumn,epjc3]{svjour3}
\usepackage[numbers,sort&compress]{natbib}
\RequirePackage[T1]{fontenc}
\usepackage{graphicx,amsmath,amsfonts,amssymb}
\usepackage{bm}
\usepackage{supertabular}
\usepackage{slashed}
\journalname{Eur. Phys. J. A}
\allowdisplaybreaks
\def\s0{\hspace{-0.20cm} & \hspace{-0.20cm}}

\begin{document}

\title{Parton distribution functions and fragmentation functions of spin-1 hadrons}    
\author{S.~Kumano\thanksref{addr1,addr2}}
\institute{
Department of Mathematics, Physics, and Computer Science,
         Faculty of Science, Japan Women's University,\\
         Mejirodai 2-8-1, Tokyo, 112-8681, Japan \label{addr1}
\and
Theory Center, Institute of Particle and Nuclear Studies,
         High Energy Accelerator Research Organization (KEK),\\
         Oho 1-1, Tsukuba, Ibaraki, 305-0801, Japan  \label{addr2}
}
\date{August 22, 2024}
\maketitle
\begin{abstract}
Structure functions of the spin-1 deuteron will be investigated 
experimentally from the late 2020's at various facilities 
such as Thomas Jefferson National Accelerator Facility,
Fermi National Accelerator Laboratory,
nuclotron-based ion collider facility,
and electron-ion colliders.
We expect that a new high-energy spin-physics field could be created 
by these projects. In this paper, the current theoretical status 
is explained for the structure functions of spin-1 hadrons,
especially on parton distribution functions,
transverse-momentum dependent parton distributions,
and fragmentation functions.
Related multiparton distribution functions are also shown.
\end{abstract}

\vspace{-0.30cm}
\section{Introduction}
\label{introduction}
\vspace{-0.15cm}

Structure functions of the spin-1/2 nucleons have been investigated
for a long time including the polarized ones. The origin of the nucleon
spin has not been identified yet because the gluon-spin and 
partonic orbital-angular-momentum contributions are not determined yet.
It should be clarified by the project of the electron-ion colliders 
(EICs) \cite{eic-2022,EicC-2021} in 2030's. 
In contrast, studies on structure functions of 
spin-1 hadrons are at a premature stage. 
It is because there is no experimental measurement
on polarized structure functions of spin-1 hadrons
since the HERMES $b_1$ experiment in 2005
\cite{hermes-b1-2005}.
We expect that the situation will change significantly 
in a few years because of various experimental projects 
on spin-1 structure functions,
such as at the Thomas Jefferson National Accelerator Facility (JLab)
\cite{Jlab-b1,jlab-gluon-trans},
the Fermi National Accelerator Laboratory (Fermilab)
\cite{Fermilab-spin,Keller-2022},
the nuclotron-based ion collider facility (NICA)
\cite{NICA-2021}, 
and the EICs.
It is likely that a new field of high-energy spin physics will be 
created for spin-1 hadrons such as the deuteron.

\begin{figure}[t!]
 \vspace{-0.00cm}
\begin{center}
   \includegraphics[width=8.3cm]{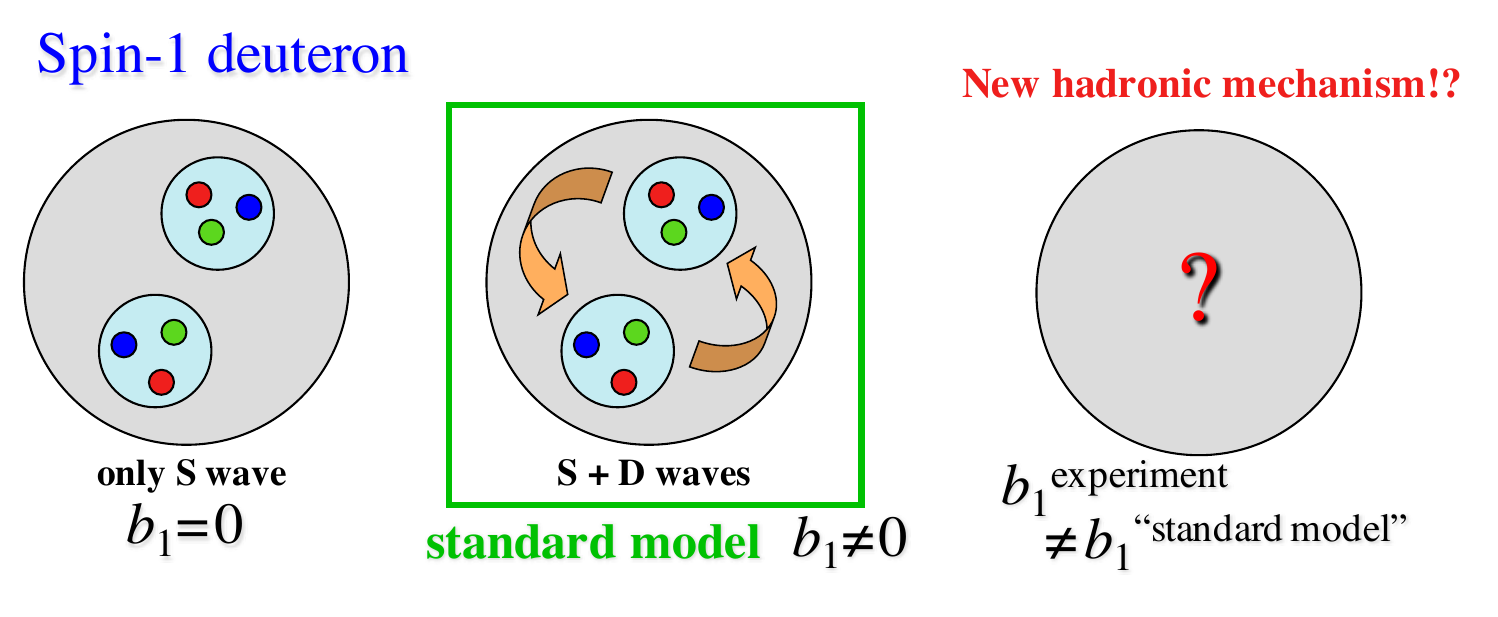}
\end{center}
\vspace{-0.5cm}
\caption{New hadron physics by structure function $b_1$.}
\label{fig:b1-new-hadron}
\vspace{-0.45cm}
\end{figure}

The tensor-polarized leading-twist structure function $b_1$ 
does not exist in the spin-1/2 nucleons 
and it is unique in a spin-1 project
\cite{fs83,hjm89}. 
If the spin-1 deuteron is an S-wave bound state of a proton 
and a neutron, $b_1$ should vanish ($b_1=0$). Since there 
is a standard deuteron model with the D-state admixture, 
we can calculate $b_1$ theoretically by the convolution model. 
However, such a ``standard-model'' calculation
\cite{b1-convolution-2017} indicated that the theoretical 
$b_1$ distribution could be very different from the HERMES 
measurement \cite{hermes-b1-2005}. 
It possibly suggests that a new hadronic mechanism
would be needed for explaining $b_1$ as illustrated
in Fig.\,\ref{fig:b1-new-hadron}.

\begin{figure}[t!]
 \vspace{-0.00cm}
\begin{center}
   \includegraphics[width=8.3cm]{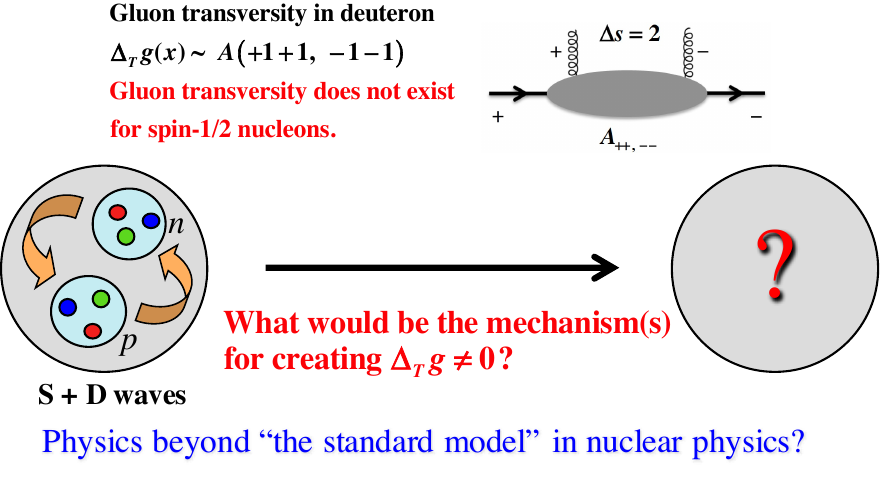}
\end{center}
\vspace{-0.5cm}
\caption{New hadron physics by gluon transversity $\Delta_T g$.}
\label{fig:g-transversity-new-hadron}
\vspace{-0.30cm}
\end{figure}

Another interesting observable is the gluon transversity $\Delta_T g$
\cite{JM-g-transversity-1989},
which does not exist in the spin-1/2 nucleons, 
because the change of two spin units is necessary for the helicity flip
amplitude as shown in Fig.\,\ref{fig:g-transversity-new-hadron}.
Since there is no contribution from a proton and a neutron
in the deuteron, if a finite $\Delta_T g$ is found in experiments 
\cite{jlab-gluon-trans,Keller-2022,NICA-2021,eic-2022,EicC-2021}, 
it means a new hadronic physics beyond
the simple bound system of the nucleons.
By the measurements of unique spin-1 observables, such as
$b_1$ and $\Delta_T g$, 
we expect that a new hadron-physics field could be created.

In order to prepare for the future experimental projects
on spin-1 hadrons, the theoretical status is explained for possible 
parton distribution functions (PDFs),
transverse-momentum-dependent parton distributions (TMDs), 
and fragmentation functions (FFs) of spin-1 hadrons in this paper.
Related multiparton distribution functions are also shown.
This article consists of the following.
In Sec.\,\ref{spin-1-sfs}, the structure functions $b_{1-4}(x)$
and the tensor-polarized PDFs are explained.
The TMDs for spin-1 hadrons are discussed up to twist 4 
in Sec.\,\ref{tmds}, 
possible fragmentation functions are shown also up to twist 4 
in Sec.\,\ref{fragmentation}, 
and the summary is given in Sec.\,\ref{summary}.

\vspace{-0.30cm}
\section{Structure functions for spin-1 hadrons}
\label{spin-1-sfs}
\vspace{-0.15cm}

Deep inelastic charged-lepton scattering from a spin-1 hadron
is described by the variables $x$ and $Q^2$, where 
$Q^2$ is given by the momentum transfer $q$ as $Q^2 = - q^2$
and the Bjorken scaling variable $x$ is by $x=Q^2 /2 M_N q^0$ 
with the nucleon mass $M_N$ and the energy transfer $q^0$.
On the other hand, the scaling variable could be defined as
$x_{\text{\tiny $D$}}=Q^2 /(2 p \cdot q)$ with the hadron momentum $p$.
In the deuteron's structure functions, for example,
the kinematical range  of $x$ is given by $0 \le x \le 2$ 
instead of $0 \le x_{\text{\tiny $D$}} \le 1$ 
because $x$ is defined by the nucleon mass.
It is sometimes confusing in handling nuclear structure functions.
The hadron tensor for the charged-lepton deep inelastic scattering 
for a spin-1 hadron is expressed in terms of eight structure functions,
$F_{1,2}$, $g_{1,2}$, and $b_{1-4}$ as
\cite{fs83,hjm89,spin-1-projection-2008,tensor-summary-2014}
\begin{align}
W_{\mu \nu}^{\lambda_f \lambda_i}
   = & -F_1 \hat{g}_{\mu \nu} 
     +\frac{F_2}{M \nu} \hat{p}_\mu \hat{p}_\nu 
     + \frac{ig_1}{\nu}\epsilon_{\mu \nu \lambda \sigma} q^\lambda s^\sigma  
\nonumber\\[-0.05cm]
&     +\frac{i g_2}{M \nu ^2}\epsilon_{\mu \nu \lambda \sigma} 
      q^\lambda (p \cdot q s^\sigma - s \cdot q p^\sigma )
\nonumber\\[-0.05cm]
& 
     -b_1 r_{\mu \nu} 
     + \frac{1}{6} b_2 (s_{\mu \nu} +t_{\mu \nu} +u_{\mu \nu}) 
\nonumber\\[-0.05cm]
&
     + \frac{1}{2} b_3 (s_{\mu \nu} -u_{\mu \nu}) 
     + \frac{1}{2} b_4 (s_{\mu \nu} -t_{\mu \nu}) ,
\label{eqn:w-1}
\end{align}
where 
$r_{\mu \nu}$, $s_{\mu \nu}$, $t_{\mu \nu}$, and $u_{\mu \nu}$
are given by
\begin{align}
r_{\mu \nu} & = \frac{1}{\nu ^2}
   \bigg [ q \cdot E ^* (\lambda_f) q \cdot E (\lambda_i) 
           - \frac{1}{3} \nu ^2  \kappa \bigg ]
   \hat{g}_{\mu \nu}, 
\nonumber\\
s_{\mu \nu} & = \frac{2}{\nu ^2} 
   \bigg [ q \cdot E ^* (\lambda_f) q \cdot E (\lambda_i) 
           - \frac{1}{3} \nu ^2  \kappa \bigg ]
\frac{\hat{p}_\mu \hat{p}_\nu}{M \nu}, 
\notag \\
t_{\mu \nu} & = \frac{1}{2 \nu ^2}
   \bigg [ q \cdot E ^* (\lambda_f) 
           \left\{ \hat{p}_\mu \hat E_\nu (\lambda_i) 
                 + \hat{p} _\nu \hat E_\mu (\lambda_i) \right\}
\notag \\[-0.10cm]
&   + \left\{ \hat{p}_\mu \hat E_\nu^* (\lambda_f)  
           + \hat{p}_\nu \hat E_\mu^* (\lambda_f) \right\}  
     q \cdot E (\lambda_i) 
   - \frac{4 \nu}{3 M}  \hat{p}_\mu \hat{p}_\nu \bigg ] ,
\notag \\
u_{\mu \nu} & = \frac{M}{\nu} 
   \bigg [ \hat E_\mu^* (\lambda_f) \hat E_\nu (\lambda_i) 
          +\hat E_\nu^* (\lambda_f) \hat E_\mu (\lambda_i) 
\notag \\[-0.10cm]
& \hspace{1.00cm} 
   +\frac{2}{3}  \hat{g}_{\mu \nu}
   -\frac{2}{3 M^2} \hat{p}_\mu \hat{p}_\nu \bigg ] .
\end{align}
Here, $\hat{g}_{\mu \nu}$ and $\hat{a}_\mu$ are defined by
$\hat{g}_{\mu \nu} \equiv  g_{\mu \nu} - {q_\mu q_\nu}/{q^2}$ and
$\hat{a}_\mu \equiv a_\mu - ({a \cdot q}/{q^2}) q_\mu $
so as to ensure the current conservation 
$q^\mu W _{\mu \nu} = q^\nu W _{\mu \nu}=0$,
$\epsilon_{\mu \nu \lambda \sigma}$ is an antisymmetric tensor
with the convention $\epsilon_{0123}=+1$, $\nu$ is defined by
$\nu ={p \cdot q}/{M}$ with the spin-1 hadron mass $M$, 
$\kappa$ is defined by $\kappa= 1+{Q^2}/{\nu^2}$, 
and $s^\mu$ is the spin vector of the spin-one hadron.
The initial and final spin states are denoted as $\lambda_i$
and $\lambda_f$, respectively, and
off-diagonal terms with $\lambda_f \ne \lambda_i$ are needed for
higher-twist terms \cite{jm89}.
The $E^\mu$ is the polarization vector of the spin-one hadron 
\vspace{-0.10cm}
\begin{align}
E^\mu (\lambda= \pm 1) \! = \! \frac{1}{\sqrt{2}}(0,\mp 1, -i,0),
\
  E^\mu (\lambda=0) \! = \! (0,0,0,1) ,
\end{align}
\ \vspace{-0.45cm} \  \\ \noindent
and it satisfies the conditions, $p \cdot E =0$ and $E^* \cdot E =-1$.
The spin vector $s$ is given by the polarization vector as
\vspace{-0.35cm}
\begin{equation}   
(s_{\lambda_f \lambda_i})^{\mu}
      = -\frac{i}{M} \epsilon ^{\mu \nu \alpha \beta} 
                E^*_\nu (\lambda_f) E_\alpha (\lambda_i) p_\beta .
\end{equation}
The structure functions $b_1$ and $b_2$ are twist-2 functions
and they satisfy the Callan-Gross type relation
$2 x_{\text{\tiny $D$}} b_1 =b_2$ 
in the Bjorken-scaling limit.
The functions $b_3$ and $b_4$ are higher-twist ones.
The projection operators for $F_{1,2}$, $g_{1,2}$, and $b_{1-4}$
from the hadron tensor $W_{\mu \nu}^{\lambda_f \lambda_i}$
are shown in Ref.\,\cite{spin-1-projection-2008}.

In the parton model, the structure function $b_1$ is written 
in terms of the tensor-polarized distributions $\delta_{_T} q (x)$ as
\cite{hjm89,b1sum}
\vspace{-0.10cm}
\begin{align}
b_1 (x,Q^2) & = \frac{1}{2} \sum_i e_i^2 
      \, \left [ \delta_{_T} q_i (x,Q^2) 
      + \delta_{_T} \bar q_i (x,Q^2)   \right ] , 
\nonumber \\[-0.10cm]
       \delta_{_T} q_i &  \equiv q_i^0 - \frac{q_i^{+1}+q_i^{-1}}{2} ,
\label{eqn:b1-parton}
\\[-0.65cm]\nonumber
\end{align}
where $i$ is the quark flavor, $e_i$ is its charge,
and $q_i^\lambda$ indicates an unpolarized-quark distribution
in the hadron with the spin state $\lambda$. 
We may note that $\delta_{_T} q (x)$ is the unpolarized quark 
distribution in a tensor-polarized hadron.
In this equation, the overall factor 1/2 is introduced in $b_1$
in the same way with $F_1$ and $g_1$ in terms of the corresponding PDFs.
There are various conventions for the tensor-polarized distribution
$\delta_{_T} q$, and it is also denoted as $\delta q $
or $f_{1LL} \, [= - (2/3) \delta_T q ]$.

\vspace{-0.30cm}
\subsection{Sum rule for $b_1$}
\label{sum-rule}
\vspace{-0.15cm}

\begin{figure}[b!]
 \vspace{-0.30cm}
   \includegraphics[width=8.0cm]{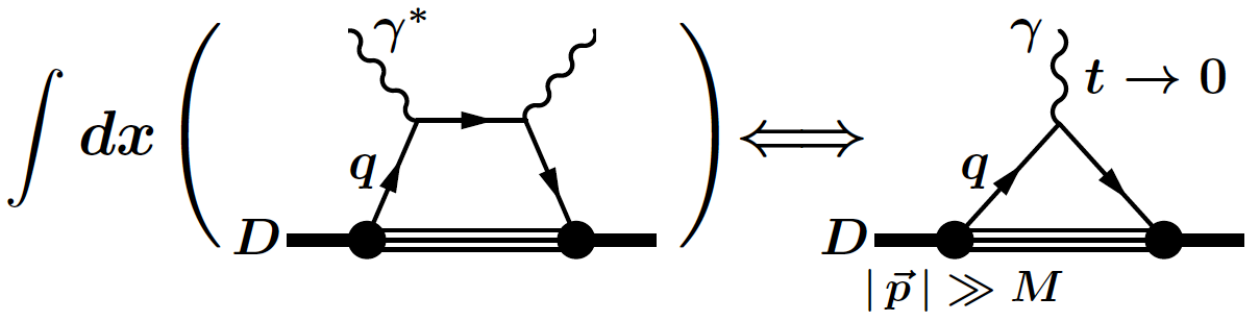}
\vspace{-0.1cm}
\caption{Relation between the first moment of a structure function
and an elastic form factor.}
\label{fig:b1-sum-fig}
\vspace{-0.00cm}
\end{figure}

Sum rules based on the parton model can be derived by relating 
a structure function integrated over $x$ to an elastic form factor 
in the infinite momentum frame \cite{feynman-book}
as shown in Fig.\,\ref{fig:b1-sum-fig}, and 
a useful sum rule was derived for $b_1$ by using this method
\cite{b1sum}.
It is not a rigorous sum rule, and it is intended to supply a basic idea 
on $b_1$  and tensor-polarized PDFs in the same way with 
the Gottfried sum rule 
\cite{KUMANO1998183,GARVEY2001203,PENG201443}.

A rough sum-rule form could be guessed in an intuitive way
by the dimensional counting.
In the hadron tensor of Eq.\,(\ref{eqn:w-1}),
$b_1$ is defined as a dimensionless quantity, so that
its first moment also does not have mass dimension.
Its sum should be expressed by a global electromagnetic observable
to satisfy the parity and time-reversal invariances, 
namely by the electric quadrupole moment $Q_h$
of the hadron $h$, for a tensor-polarized spin-1 hadron.
The $Q_h$ has the dimension of length$^2 =$1/mass$^2$, 
so that it should be cancelled by a mass$^2$ factor as
$\int dx b_1 (x) = \text{(mass dim.)}^2 Q_h$.
From the elastic process in Fig.\,\ref{fig:b1-sum-fig},
this mass-dimension squared factor could be either 
the hadron mass squared $M^2$
or the momentum transfer squared $t$.

By the optical theorem, the hadron tensor is equal to 
the imaginary part of the forward scattering amplitude 
for virtual photon scattering. It indicates that
the first moment of a structure function is related to
an elastic form factor expressed by the PDFs
in the infinite-momentum frame 
as illustrated in Fig.\,\ref{fig:b1-sum-fig}.
For example, the polarized Bjorken sum rule 
was derived by this method in Ref.\,\cite{feynman-book}.
This method was applied for $b_1$ of the deuteron \cite{b1sum}.
Here, the structure function $b_1$ and the PDFs are defined by
the ones per nucleon and the scaling variable $x$ is used.
As illustrated in the left-hand side of Fig.\,\ref{fig:b1-sum-fig},
integrating Eq.\,(\ref{eqn:b1-parton}) over $x$
and using the relations
$ ( \delta_{_T} u_v )_D \equiv ( \delta_{_T} u - \delta_{_T} \bar u )_D
                  = (\delta_{_T} u_v^p + \delta_{_T} u_v^n)/2
                  = (\delta_{_T} u_v + \delta_{_T} d_v)/2$,
$ ( \delta_{_T} d_v )_D = (\delta_{_T} d_v + \delta_{_T} u_v)/2$         
for the valence-quark distributions, we obtain
\vspace{-0.10cm}
\begin{align}
\int dx \, b_1 (x) 
   &   = \frac{5}{36} \int dx
      \, \left [ \, \delta_{_T} u_v (x) + \delta_{_T} d_v (x) \, \right ]
\nonumber \\[-0.10cm]
   & \ \ 
      + \sum_i e_i^2 \int dx \, \delta_T \bar q_{i,D} (x) ,
\label{eqn:b1-sum-parton-1}
\end{align}
\ \vspace{-0.45cm} \  \\ \noindent
where $Q^2$ dependence is not explicitly shown and
the relation 
$ \int dx (\delta_T s - \delta_T \bar s )_D
 =\int dx (\delta_T c - \delta_T \bar c )_D
 =\int dx (\delta_T b $ $- \delta_T \bar b )_D = 0$
is used.
The function $q_v$ indicates a valence-quark distribution,
and the valence-quark distributions in the deuteron 
come from the corresponding ones in the proton and the neutron.

The helicity amplitude with the charge operator $J_0$ 
is defined for the elastic scattering
in the right-hand side of Fig.\,\ref{fig:b1-sum-fig} .
If the frame with a large longitudinal momentum $ | \, \vec p \, | \gg M$ 
is taken, the elastic amplitude is calculated by the parton model as
\cite{feynman-book}
\begin{align}
\Gamma_{H,H} & \equiv \langle \, p,  H \, \left | \, J_0 (0) 
                     \, \right | \, p, H \, \rangle 
\nonumber \\
 & = \sum_i e_i \int \! dx
\left [ q_i^H (x) - \bar q_i ^H (x) \right ]_D .
\label{eqn:hecility-amp-1}
\end{align}
\ \vspace{-0.45cm} \  \\ \noindent
The tensor combination of the amplitudes is expressed
in terms of the valence-quark distributions in the deuteron 
by considering the proton and neutron contributions as
$
\Gamma_{0,0} - ({\Gamma_{1,1} + \Gamma_{-1,-1} })/{2}
= \frac{1}{3} \int dx \, \left [ \, \delta_{_T} u_v (x) 
                                  + \delta_{_T} d_v (x)   \right ]
$
with the condition 
$\int dx ( \delta_{_T} s - \delta_{_T} \bar s )_D = 
 \int dx ( \delta_{_T} c - \delta_{_T} \bar c )_D = 
 \int dx ( \delta_{_T} b - \delta_{_T} \bar b )_D = 0$.
Substituting this helicity relation into 
Eq. (\ref{eqn:b1-sum-parton-1}), we obtain 
\begin{align}
\int dx \, b_1 (x) 
     & = \frac{5}{12}
\, \left [ \, \Gamma_{0,0} - \frac{ \Gamma_{1,1}  + \Gamma_{-1,-1}}{2} \, 
   \right  ]
\nonumber \\[-0.10cm]
& \ \ \,
      + \sum_i e_i^2 \int dx \, \delta_T \bar q_{i,\text{\tiny $D$}} (x) .
\label{eqn:b1-sum-hecility}
\end{align}
\vspace{-0.40cm}

The helicity amplitudes  are expressed by 
the electric charge and quadrupole form factors of the deuteron,
$F_C (t)$ and $F_Q (t)$
expressed by the momentum-transfer squared $t$ ($\to 0$),
as
\begin{align}
\Gamma_{0,0} & = 
\lim_{t \to 0} 
\left [ F_C (t) - \frac{t}{3 M^2} F_Q (t)  \right ] , 
\nonumber \\
\Gamma_{1,1} & = \Gamma_{-1,-1} = 
\lim_{t \to 0} 
     \left [ F_C (t) + \frac{t}{6 M^2} F_Q (t)  \right ] ,
\label{eqn:form-factor-1}
\end{align}
where $F_C (t)$ and $F_Q (t)$ are defined by the units 
of $e$ and $e/M^2$, respectively, with the elementary charge $e$ 
and the deuteron mass $M$.
The tensor-polarization combination of the helicity amplitudes
is given by the quadrupole form factor as
$
\Gamma_{0,0} - (\Gamma_{1,1} + \Gamma_{-1,-1} )/2
= - \lim_{t \to 0} \frac{t}{2 M^2} F_Q (t)$ $ = 0 $.
Using this relation, we obtain the sum rule
\begin{align}
\int \! dx \, b_1 (x) 
      = - \lim_{t \to 0} \frac{5 \, t}{24 M^2} F_Q (t)
   + \! \sum_i e_i^2 \! \int \! dx \, \delta_T \bar q_{i,\text{\tiny $D$}} (x) .
\label{eqn:b1-sum-parton-3}
\end{align}
\ \vspace{-0.45cm} \  \\ \noindent
If the tensor-polarized antiquark distributions vanish
$ \int dx \, \delta_T \bar q_{i,\text{\tiny $D$}} (x) = 0$,
the sum becomes
\begin{align}
\int dx \, b_1 (x) 
      = 0 .
\label{eqn:b1-sum-parton-4}
\end{align}
\ \vspace{-0.75cm} \  \\ \noindent

This $b_1$ sum rule is similar to
the Gottfried sum rule which was also derived by the parton model:
\begin{alignat}{2}
\! \! \!
\int dx \, b_1 (x) 
   &  =  0 &
   &  + \sum_i e_i^2 \int \! dx \, \delta_T \bar q_{i,\text{\tiny $D$}} (x) ,
\nonumber \\
\! \! \!
\int \frac{dx}{x} \, [F_2^p (x) - F_2^n (x) ]
   &  = \frac{1}{3} &
   &  +  \frac{2}{3} \int \! dx \, [ \bar u(x) - \bar d(x) ] .
\label{eqn:b1-sum-gottfried}
\end{alignat}
The finite Gottfried sum 1/3 is due to
the flavor dependence of the valence quark distributions,
$\int [ u_v(x) - d_v (x) ]/3=1/3$.
On the other hand, the $b_1$ sum vanishes 
($ -  \lim_{t \to 0} (5/24) t F_Q (t) =0$)
because the valence-quark number does not depend 
on the tensor polarization.
The second moment of $b_1$ was also shown to vanish $\int dx x b_1(x)=0$
\cite{ET-1982}, which could be related to an interesting
shear-force property \cite{NICA-2021}.

We may note that both sum rules are not rigorous ones 
and they are based on the parton model.
The integrals
$\int \! dx \, \delta_T \bar q_{i,D} (x)$ and
$\int \! dx \, [ \bar u(x) - \bar d(x) ]$ 
could diverge depending on the functional form 
of $x$ at small $x$, especially as $Q^2$ increases.
This was studied in Ref.\,\cite{miller-2014}
in a convolution model.
Although both sum rules may not be satisfied especially
at large $Q^2$, we know, nonetheless, that
the Gottfried sum rule was very useful 
for indicating the difference between $\bar u$ and $\bar d$
from its violation in DIS experiments, and
such studies created a new research line in hadron physics on 
flavor asymmetric antiquark distributions ($\bar u - \bar d$) 
in the nucleon
\cite{KUMANO1998183,GARVEY2001203,PENG201443}.

In the same way, the finite sum of $b_1$ could indicate
the tensor-polarized antiquark distributions
$ \delta_T \bar q_{i,\text{\tiny $D$}} (x) $,
and the sum for the valence-quark part ($\int b_{1,\text{val}} =0$) 
provides a guideline 
on the functional form of the tensor-polarized 
valence-quark distributions as used, for example,
in Sec.\,\ref{parametrization}.
About the antiquark distributions, there was already a hint 
from the HERMES measurement as \cite{hermes-b1-2005}
$
\int_{0.02}^{0.85} dx b_1(x)  =
   [0.35 $ $ \pm 0.10$ $ \text{ (stat)} \pm 0.18 \text{ (sys)}] \times 10^{-2} 
$
at $Q^2>1 \, \text{GeV}^2$.
If the JLab experiment finds a finite sum with a reasonable precision
\cite{Jlab-b1}, the mechanism of creating a finite 
$ \delta_T \bar q_{i,\text{\tiny $D$}} (x) $
will become an interesting theoretical topic.

\vspace{-0.30cm}
\subsection{Parametrization for tensor-polarized PDFs}
\label{parametrization}
\vspace{-0.15cm}

The HERMES data \cite{hermes-b1-2005}
are the only available ones for the structure function $b_1$, 
so that a real global analysis cannot be done 
for the tensor-polarized PDFs at this stage.
However, it is useful to provide appropriate tensor-polarized PDFs 
for planning future experiments and 
for testing theoretical model and possible lattice-QCD calculations.
We define the PDF parametrization form as
\cite{tensor-pdfs}
\begin{align}
\delta_{_T} q_{iv}^{\text{\tiny $D$}} (x) 
  & = \delta_{_T} w(x) \, q_{iv}^{\text{\tiny $D$}} (x), 
\nonumber \\ 
\delta_{_T} \bar q_i^{\text{\tiny $D$}} (x) 
  & = \alpha_{\bar q} \, \delta_{_T} w(x) \, \bar q_i^{\text{\tiny $D$}} (x) ,
\label{eqn:delta-q-qbar-1}
\end{align}
\ \vspace{-0.35cm} \  \\ \noindent
at the initial scale $Q_0^2$
by considering that certain fractions of the unpolarized quarks
are tensor polarized. At this stage, there is no data to indicate
the tensor-polarized gluon distribution $\delta_T g$,
so that $\delta_T g =0$ is assumed at $Q_0^2$.
It means that a finite $\delta_T g$ exists only  at a different scale $Q^2$
as shown in Ref.\,\cite{pd-Drell-Yan-tensor-2016}.
We assumed the common function $\delta_{_T} w(x)$ for both
the valence-quark and antiquark distributions.
The $\alpha_{\bar q}$ is a parameter for the difference
between the overall valence-quark and antiquark distributions,
and its flavor dependence was not considered.
Neglecting small nuclear corrections, of the order of a few percent
in the deuteron, we write the PDFs of the deuteron as 
$q_i^D=(q_i^{\,p}+q_i^{\,n})/2$ and 
$\bar q_i^D=(\bar q_i^{\,p}+\bar q_i^{\,n})/2$.
Then, the isospin symmetry is used for the PDFs in the neutron
and flavor symmetric tensor-polarized distributions are defined as
\begin{align}
\delta_{_T} q_v^{\text{\tiny $D$}} (x)
        & \equiv \delta_{_T} u_v^{\text{\tiny $D$}} (x) 
         = \delta_{_T} d_v^{\text{\tiny $D$}} (x) 
         = \delta_{_T} w(x) \, \frac{u_v (x) +d_v (x)}{2} , 
\nonumber \\
\delta_{_T} \bar q^{\text{\tiny $D$}} (x)
        & \equiv \delta_{_T} \bar u^{\text{\tiny $D$}} (x)
         = \delta_{_T} \bar d^{\text{\tiny $D$}} (x)
         = \delta_{_T}      s^{\text{\tiny $D$}} (x)
         = \delta_{_T} \bar s^{\text{\tiny $D$}} (x)                    
\nonumber \\
& \! \! \!
    = \alpha_{\bar q} \, \delta_{_T} w(x) \, 
    \frac{2 \bar u(x) +2 \bar d(x) +s(x) + \bar s(x)}{6} .
\label{eqn:dw(x)}
\end{align}
The tensor-polarized charm and bottom quark distributions are neglected.

The leading-order (LO) expression for $b_1$  of Eq.\,(\ref{eqn:b1-parton})
is given by these PDFs as
\begin{align}
b_1^{\text{\tiny $D$}} (x) & = \frac{1}{36} \delta_{_T} w(x) \, \big [ \,
     5 \{ u_v (x) + d_v (x) \}   
\nonumber \\
& \ \ \ \ 
  +4 \alpha_{\bar q} \{ 2 \bar u (x) + 2 \bar d (x)
  +   s (x) + \bar s (x) \} \, \big ] .
\label{eqn:b1x}
\end{align}
Because there is no experimental indication for the scaling violation,
the $Q^2$ dependence is ignored. 
The scale $Q_0^2$ is taken as the average one $Q_0^2=2.5$ GeV$^2$
of the HERMES experimental values.
As shown in Sec.\,\ref{sum-rule}, 
the tensor-polarized valence-quark distributions satisfy 
the relation $\int dx (b_1)_{\text{valence}}=0$, so that 
the weight function $\delta_{_T} w(x)$ could be parametrized as
\begin{equation}
\delta_{_T} w(x) = a x^b (1-x)^c (x_0-x) ,
\label{eqn:dw(x)-abc}
\end{equation}
with the parameters $a$, $b$, and $c$, which are determined
by a $\chi^2$ analysis.
The node factor $x_0$ is expressed in terms of these parameters 
by the condition $\int dx (b_1)_{\text{valence}}$ $=0$.
For satisfying this integral, a node should exist in the tensor-polarized
PDFs and it is also supported by the convolution model
\cite{b1-convolution-2017}.

\begin{figure}[t!]
   \begin{center}
   \vspace{0.0cm}
   \includegraphics[width=6.0cm]{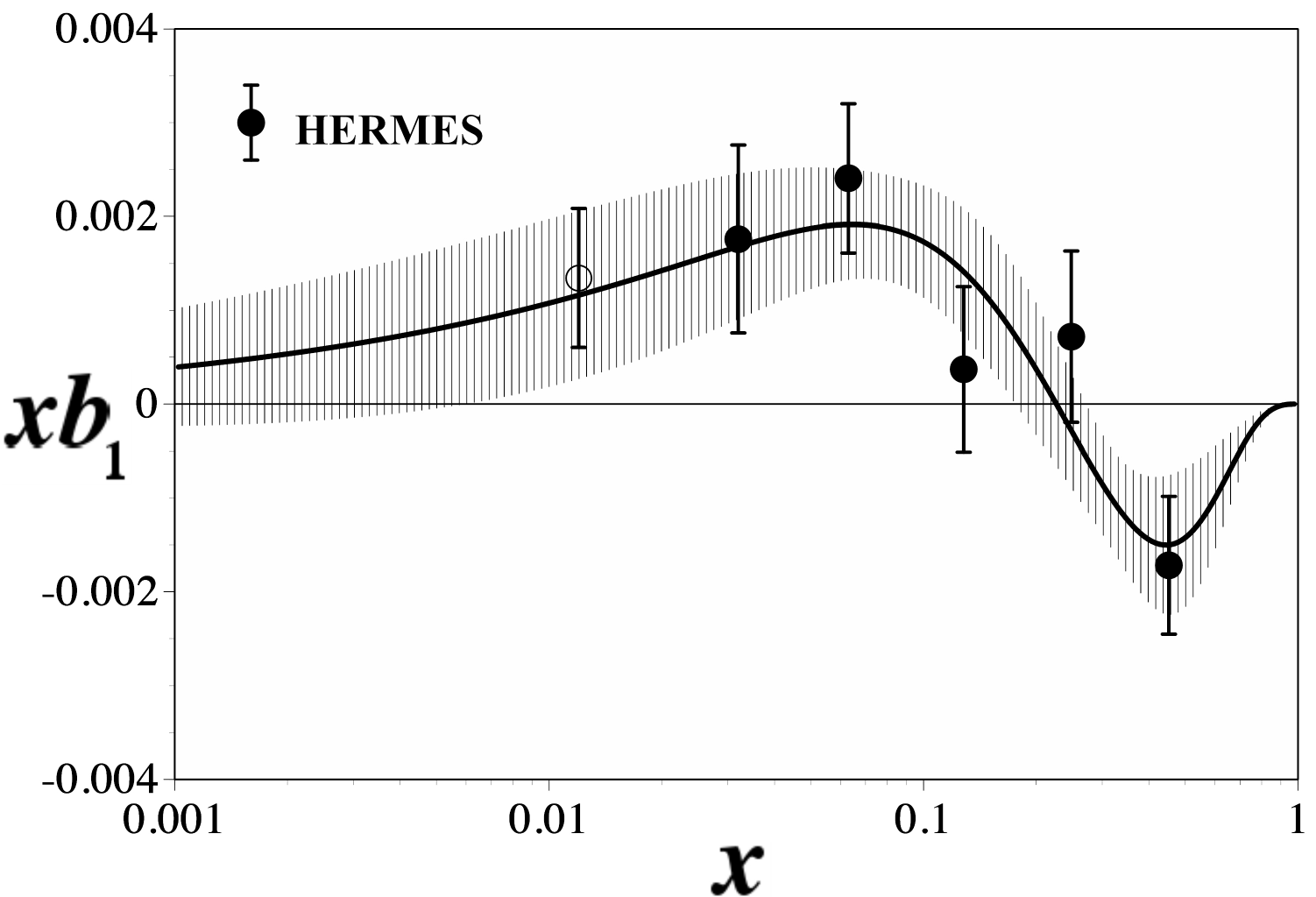}
   \end{center}
\vspace{-0.3cm}
\caption{HERMES data and parametrized $b_1$ structure function 
with its uncertainty band given by $\Delta \chi^2=1$.
The data with $Q^2<1$ GeV$^2$ is shown by the open circle.}
\label{fig:hermes-data}
\vspace{0.20cm}
   \begin{center}
   \includegraphics[width=6.0cm]{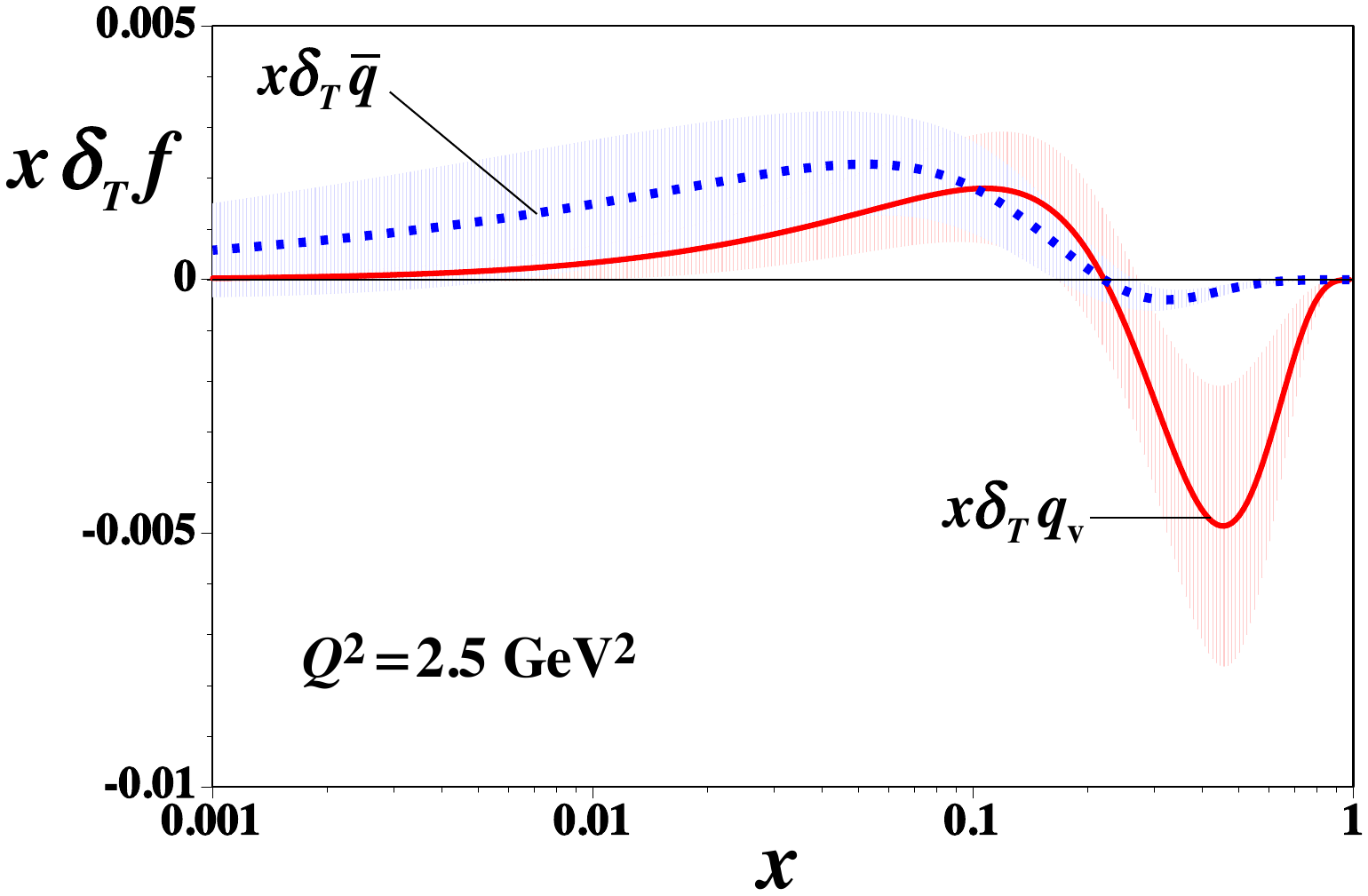}
   \end{center}
\vspace{-0.3cm}
\caption{Determined tensor polarized distributions are shown 
with their uncertainties given by $\Delta \chi^2=1$.}
\label{fig:tensor-distribution}
\end{figure}

The parameters were determined by analyzing the HERMES data 
with ($\alpha_{\bar q} \ne 0$) or without  ($\alpha_{\bar q} = 0$)
the tensor-polarized antiquark distributions in Ref.\,\cite{tensor-pdfs}
and the optimum fit was with $\alpha_{\bar q} \ne 0$ (set 2).
The parametrized $b_1$ is calculated with the determined 
parameters and it is compared with the HERMES data 
in Fig.\,\ref{fig:hermes-data}
\cite{tensor-pdfs,pd-Drell-Yan-tensor-2016}.
The obtained tensor-polarized PDFs are shown in Fig.\,\ref{fig:tensor-distribution}
with their uncertainties obtained by $\Delta \chi^2=1$.
The solid and dotted curves are
the tensor-polarized valence-quark and antiquark distributions,
respectively. Although the errors are still large,
the HERMES data seem to indicate the existence of 
finite tensor-polarized antiquark distributions.
From the parametrized $b_1$, we obtain the sum
$\int dx \, b_1 (x) = 0.0058 \pm 0.0047$, which could suggest
the finite tensor-polarized antiquark distributions according to
Eq.\,(\ref{eqn:b1-sum-gottfried}).
The JLab experiment will measure $b_1$ in the range
$0.1<x<0.5$ more accurately \cite{Jlab-b1}, 
so that we expect to obtain reliable tensor-polarized PDFs 
in the near future.
Furthermore, there are projects with the polarized deuteron target
and beam at Fermilab \cite{Fermilab-spin,Keller-2022} and NICA \cite{NICA-2021}
for investigating the tensor-polarized PDFs, for example, by
the proton-deuteron Drell-Yan process 
\cite{pd-Drell-Yan,pd-Drell-Yan-antiquark}.

\noindent
\vspace{-1.00cm}

\vspace{+0.05cm}
\subsection{Theoretical $b_1$ by the standard deuteron model}
\label{convolution}
\vspace{-0.25cm}

In order to find a new aspect of hadronic physics by the structure 
function $b_1$ in comparison with the HERMES and future data, 
$b_1$ was calculated by the standard deuteron model
in the convolution formalism \cite{b1-convolution-2017}.
For describing nuclear structure functions, the convolution model
is usually used by expressing the nuclear hadronic tensor $W^A_{\mu\nu}$ 
as a convolution of a nuclear spectral function $S(p)$ and 
the nucleonic hadronic tensor $W^N_{\mu\nu}$:
\vspace{-0.15cm}
\begin{align}
 W^A_{\mu\nu}(P_A,q)=\int d^4p \, S(p) \, W^N_{\mu\nu}(p,q) .
\end{align}
\ \vspace{-0.45cm} \ \\ \noindent
Two theoretical calculations were provided 
in Ref.\,\cite{b1-convolution-2017}.

In the first convolution method (Theory 1), scaling-limit relations 
between virtual photon-hadron helicity amplitudes 
and the structure functions of the deuteron
/nucleon are used.
Then, $b_1$ is written by the convolution integral of 
the lightcone momentum distribution of the nucleon 
$f^H(y)$, where $H$ is the deuteron polarization,
with the average of proton and neutron structure 
functions $F_1^N=(F^p_1+F^n_1)/2$ as
\begin{align}
b_1(x,Q^2) \! = \! \! \int \! \frac{dy}{y} \!
\left[f^0(y)-\frac{f^+(y)+f^-(y)}{2}\right] \!
F^N_1(x/y,Q^2) .
\label{eq:b1conv}
\end{align}
\ \vspace{-0.45cm} \ \\ \noindent
The distribution $f^H(y)$ is given by
the deuteron wave function $\phi^H(\bm p)$
with the normalization $\int d^3\bm p\,y\,|\phi^H(\bm p)|^2$ $=1$
as
\begin{equation}
\! 
f^H(y) \! = \! \! \int \!
d^3\bm p\,y\,|\phi^H(\bm p)|^2\delta \!
\left( \! y-\frac{\sqrt{m_N^2+\bm 
p^2}-p_z}{m_N} \! \right) \! .
\end{equation}
The LO expression is used for $F_1^N$ from the PDFs
by considering the transverse-longitudinal cross section ratio
$R=\sigma_L/\sigma_T$ as
\begin{align}
F^N_1(x,Q^2) & =\frac{1+4m_N^2x^2/Q^2}{2x[1+R(x,Q^2)]}
\nonumber \\
& \ \ 
\times
\sum_i e^2_i \, x \left[ q_i (x,Q^2)+\bar{q}_i (x,Q^2)\right] .
\end{align}

In the second convolution method (Theory 2), the virtual nucleon 
approximation is used. It considers the $np$ component of 
the light-front deuteron wave function, and the virtual photon 
interacts with one nucleon ($i$) which is off the mass shell, 
while the second noninteracting spectator nucleon ($N$) 
is assumed to be on its mass shell.
In this method, no scaling limit relations are used, so that
higher-twist nuclear effects are included. 
Its $b_1$ expression is given by
\begin{align}
\label{eq:b1vna}
& b_1(x,Q^2) =\frac{3}{4(1+Q^2/\nu^2)} \int \frac{k^2}{\alpha_i} 
dk \, d(\cos\theta_k)
\nonumber \\
& \ \ \ 
\times
\bigg[ F_{1}^N(x_i,Q^2) \left(6\cos^2\theta_k-2\right) 
+\frac{\bm p_i^{\perp 2} } {2 \, p_i q } F_{2 }^N (x_i ,Q^2)
\nonumber \\
& \ \ \ \ \ \ \ \ 
\times
\left(5\cos^2\theta_k-1\right) \bigg]
\left[ \frac{U(k)W(k)}{\sqrt{2}}+\frac{W( k)^2}{4}\right] ,
\end{align}
where $\nu$ is the virtual photon energy in the deuteron rest frame,
$p_i$ is the four-momentum of the struck nucleon,
$x_i=Q^2/(2 p_i \cdot q)$ is the Bjorken variable,
and $\alpha_i=2p_i^-/P^-$ is the lightcone momentum fraction
with $P=p_i+p_N$.
The momentum $k$ corresponds to the relative momentum 
of the two nucleons with identical light-front
momentum components ($P^-$, $P^\perp$).
The radial $S$- and $D$-wave light-front deuteron wave functions are denoted
as $U(k)$ and $W(k)$, and they satisfy the baryon and momentum sum rules.
The difference between both methods in Eqs.\,(\ref{eq:b1conv}) and (\ref{eq:b1vna})
comes mainly in the $F_2^N$ term which reflects the higher-twist nuclear effects.

\begin{figure}[t!]
 \vspace{-0.00cm}
 \begin{center}
   \includegraphics[width=6.0cm]{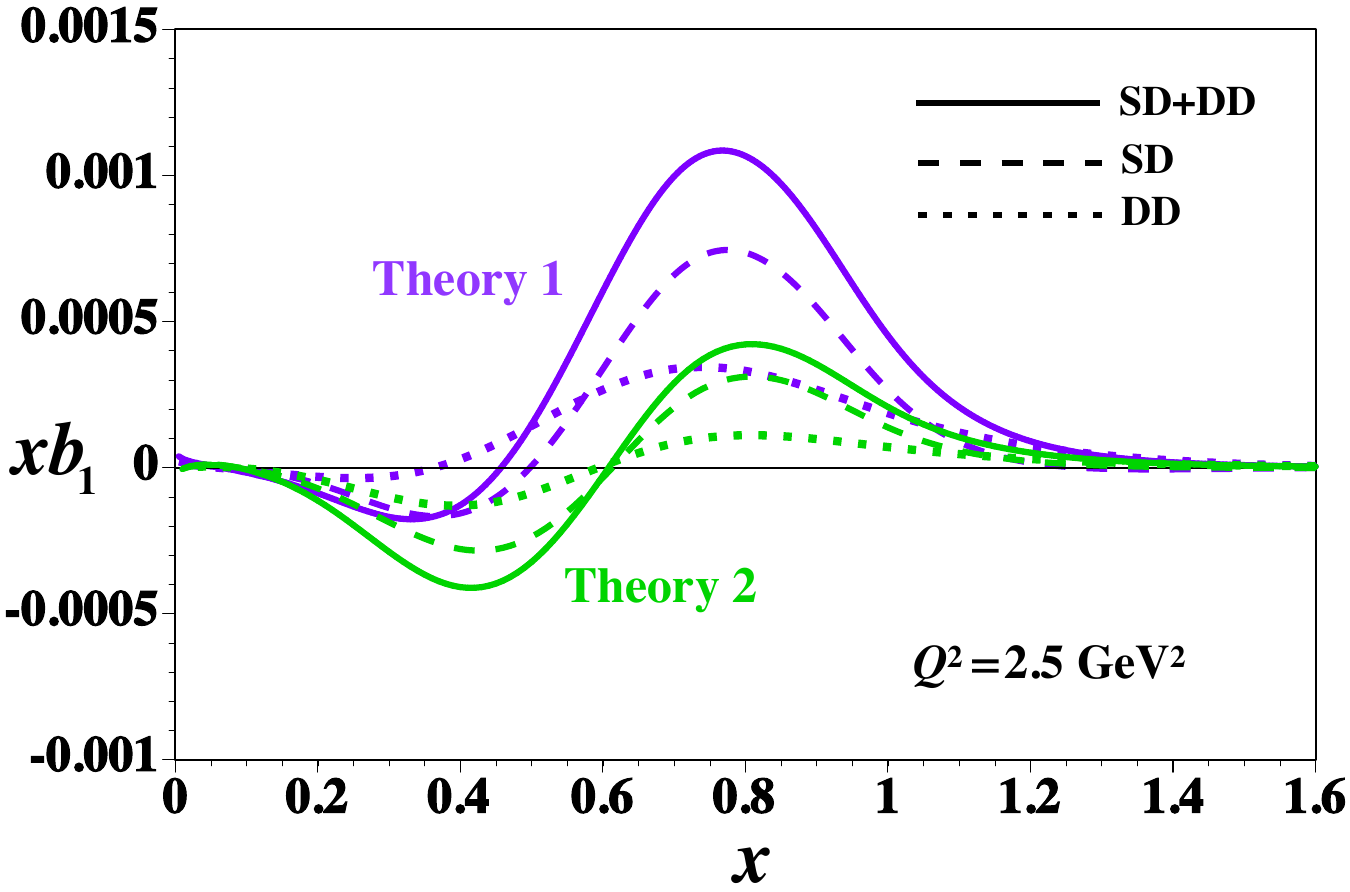}
 \end{center}
\vspace{-0.3cm}
\caption{$S$ and $D$ wave contributions to $xb_1$.
The SD and DD indicate S-D interference and
the pure D-wave terms, respectively.}
\label{fig:xb1-2_5-sdtotal}
\vspace{-0.30cm}
\end{figure}

\begin{figure}[t!]
 \vspace{+0.20cm}
  \begin{center}
   \includegraphics[width=6.0cm]{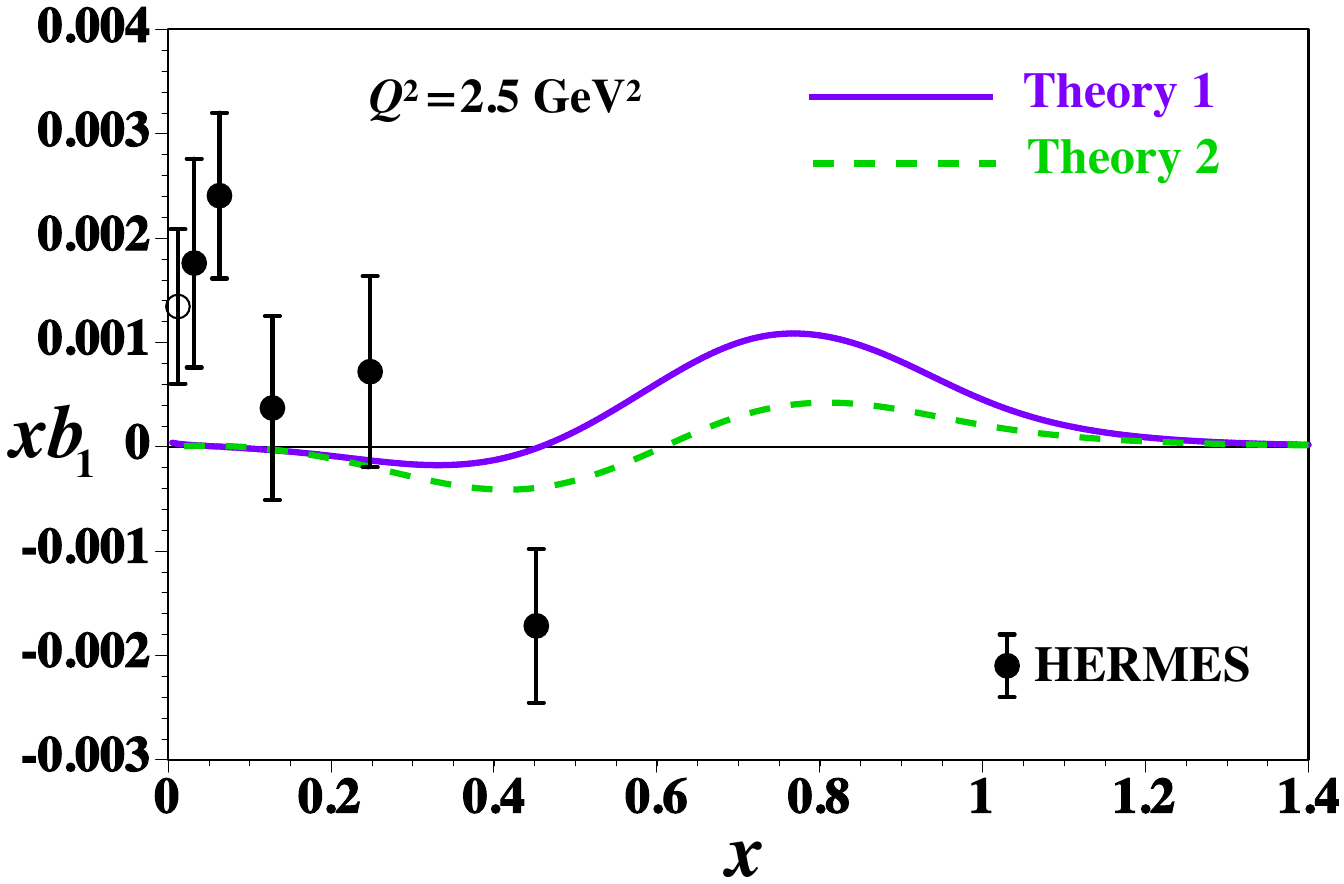}
  \end{center}
\vspace{-0.3cm}
\caption{Standard deuteron model calculations 
at $Q^2=2.5$ GeV$^2$ are compared with the HERMES data.}
\label{fig:xb1-theory-hermes}
\vspace{-0.30cm}
\end{figure}

For the numerical evaluations, we used 
the MSTW-2008 (Martin-Stirling-Thorne-Watt, 2008) LO parametrization 
for $F^N_2$, the SLAC-R1998 parametrization for $R$, 
and the CD-Bonn deuteron wave function. 
The S and D wave contributions to $b_1$ are shown
at $Q^2$=2.5 GeV$^2$ in Fig.\,\ref{fig:xb1-2_5-sdtotal}.
The S-D interference contribution is large; however,
the pure D wave term is also comparable in magnitude. 
The total contributions are compared with the HERMES data
at $Q^2$=2.5 GeV$^2$ in Fig.\,\ref{fig:xb1-theory-hermes}.
The differences between the two theoretical results should
come mainly from higher-twist effects.
It is interesting to find finite distributions even at $x>1$.
We also notice the theoretical curves are very different
from the HERMES measurements. Since our theoretical estimates
are based on the standard deuteron model, the deviations
from the experimental data may indicate a new hadron mechanism
in $b_1$, although it would be too early to conclude the existence
of such a new mechanism. In fact, the errors of the HERMES data
are large and careful studies would be necessary for estimating
the higher-twist effects in extracting $b_1$ from
experimental measurements.
On the other hand, there is an interesting suggestion that 
$b_1$ could contain a new hadronic mechanism, such as
the hidden color \cite{miller-2014}.
Since the $Q^2$ region of the JLab experiment is similar,
one could consider the $xb_1$ curves in Figs.\,\ref{fig:xb1-2_5-sdtotal}
and \ref{fig:xb1-theory-hermes} as theoretical predictions
for the JLab experiment.

\vspace{-0.30cm}
\subsection{Gluon transversity}
\label{gluon-tansversity}
\vspace{-0.15cm}

The quark transversity distributions have been investigated 
experimentally for the spin-1/2 nucleon; however, there is
no experimental measurement for the gluon transversity 
$\Delta_T g (x)$.
It is given by the gluon-helicity flip amplitude
\begin{align}
\Delta_T g (x)  \sim \text{Im} \, A_{++,\, - \hspace{0.03cm} -} ,
\end{align}
where $A_{\Lambda_i \lambda_i ,\, \Lambda_f \lambda_f}$ is
the parton-hadron forward scattering amplitude
with the initial and final hadron helicities
$\Lambda_i$ and $\Lambda_f$ and parton ones
$\lambda_i$ and $\lambda_f$.
A change of two spin units ($\Delta s=2$) is needed
for $\Delta_T g$ as illustrated in Fig.\,\ref{fig:g-transversity-new-hadron}.
This spin-flip process is not possible in the spin-1/2 nucleon; 
however, this gluon transversity exists in the spin-1 deuteron.
It is a unique project in the deuteron. 
As explained in the introduction, this situation makes
it an interesting observable to find an exotic hadron physics
mechanism. 

The gluon transversity is defined by the matrix element
between the linearly polarized ($E_x$) deuteron as 
\cite{g-tran-drell-yan-2020}
\begin{align}
\Delta_T g (x) 
& = \varepsilon_{TT}^{\alpha\beta}
\int \frac{d \xi^-}{2\pi} \, x p^+ \, e^{i x p^+ \xi^-}
\nonumber \\[-0.10cm]
& \ \hspace{0.30cm}
\times 
\langle \, p \, E_{x} \left | \, A_{\alpha} (0) \, A_{\beta} (\xi)  
\right | p \, E_{x} \, \rangle 
_{\xi^+=\vec\xi_\perp=0}  ,
\label{eqn:pdf-definitions}
\end{align}
where 
$p$ is the deuteron momentum, $A^\mu$ is the gluon field,
$x$ is the momentum fraction for a gluon, 
$\varepsilon_{TT}^{\alpha\beta}$ is the transverse antisymmetric tensor 
with $\varepsilon_{TT}^{11}=+1$ and $\varepsilon_{TT}^{22}=-1$,
and $\xi$ is the space-time coordinate expressed by the lightcone 
coordinates $\xi^\pm = (\xi^0 \pm \xi^3)/\sqrt{2}$ and $\vec\xi_\perp$.
It is written by the gluon distribution difference as
\begin{align}
\Delta_T g (x) = g_{\hat x/\hat x} (x) - g_{\hat y/\hat x} (x) ,
\label{eqn:gluon-transversity-linear}
\nonumber \\[-0.50cm]
\end{align}
with the notation $\hat y/\hat x$  which indicates 
the gluon linear polarization $\varepsilon_y$ 
in the deuteron with the linear polarization $E_x$.
Therefore, it is the difference between gluons linearly polarized 
along $y$ or $x$ in a deuteron linearly polarized along $x$.
We may note that there are various notations for 
the same gluon transversity as
$\Delta_2 G (x)$ \cite{Artru-Mekhfi-1990,Mulders-Rodrigues-2001},
$a (x)$ \cite{JM-g-transversity-1989,Nzar-Hoodbhoy-1992}, 
$\Delta_L g (x)$ \cite{Vogelsang-1998},
$-\delta G (x)$ \cite{Sather-Schmidt-1990,Detmold-Shanahan-2016},
$-h_{1TT,g} (x)$ \cite{BM-2000,Bohr-2016,Meissner-2007},
and 
$ \Delta_T g (x)$ \cite{QCD-handbook-1995,g-tran-drell-yan-2020}
used in this paper.

In order to understand the linear polarization in comparison
with the longitudinal and transverse polarizations,
we list them in terms of  the polarization vectors
and the spin-vector/tensor parameters
of a spin-1 hadron in Table \ref{table:polarizations}.
The polarization parameters ($S_{T}^x$, $S_{T}^y$, $\cdots$)
are given by the spin vector $\bm S$ and tensor $T_{ij}$
defined by the polarization vector $\bm E$ as
\cite{BM-2000}
\ \vspace{-0.00cm}
\begin{align}
& \bm S
= \text{Im} \, (\, \bm E^{\, *} \times \bm E \,)
= (S_{T}^x,\, S_{T}^y,\, S_L) ,
\nonumber \\[-0.15cm]
& T_{ij} 
 = \frac{1}{3} \delta_{ij} 
       - \text{Re} \, (\, E_i^{\, *} E_j \,) 
\nonumber \\[-0.02cm]
& \ \hspace{+0.33cm}
= \frac{1}{2} 
\left(
    \begin{array}{ccc}
     - \frac{2}{3} S_{LL} + S_{TT}^{xx}    & S_{TT}^{xy}  & S_{LT}^x  \\[+0.20cm]
     S_{TT}^{xy}  & - \frac{2}{3} S_{LL} - S_{TT}^{xx}    & S_{LT}^y  \\[+0.20cm]
     S_{LT}^x     &  S_{LT}^y              & \frac{4}{3} S_{LL}
    \end{array}
\right) .
\label{eqn:spin-1-vector-tensor}
\end{align}
The linear polarizations are expressed by
$\bm E_x  = \left ( \, 1,\, 0,\, 0 \, \right )$ and
$\bm E_y  = \left ( \, 0,\, 1,\, 0 \, \right )$,
and they contain the linear polarization parameter $S^{xx}_{TT}$.
However, they also have the tensor polarization parameter $S_{LL}$,
so that the cross section combination $d\sigma (E_x) - d\sigma (E_y)$
needs to be taken for extracting the pure linear polarization
component. We also notice that the transverse polarizations
contain the linear polarization parameter $S^{xx}_{TT}$
in addition to the vector polarization parameters $S^i_T$ and
the tensor polarization parameter $S_{LL}$.

There is a project to measure the gluon transversity at JLab
in the electron-deuteron scattering by observing 
the dependence on the angle \cite{jlab-gluon-trans,MWZ-2013}, 
which is between the lepton-scattering plane 
and the target-spin orientation.
On the other hand, it is possible to investigate the gluon transversity
at Fermilab in the SpinQuest project \cite{Keller-2022}.
The linear-polarization difference \\ $d\sigma (E_x) - d\sigma (E_y)$
for the proton-deuteron Drell-Yan process $p(A)+d(B) \to \mu^+ \mu^- +X$ 
contains the gluon transversity $\Delta_T g$ \cite{g-tran-drell-yan-2020}
and is given by
\begin{align}
& \frac{ d \sigma_{pd \to \mu^+ \mu^- X} }{d\tau \, d \bm q_T^{\, 2} \, d\phi \, dy}
(E_x-E_y )
 = - \frac{\alpha^2 \, \alpha_s \, C_F \, q_T^2}{6\pi s^3} \cos (2\phi) 
\nonumber \\
&
\times 
\int_{\text{min}(x_a)}^1 \! \! dx_a 
 \frac{ \sum_{q}  e_q^2 \, x_a \!
 \left[ \, q_A (x_a) + \bar q_A (x_a) \, \right ] x_b \Delta_T g_B (x_b)} 
 { (x_a x_b)^2 \, (x_a -x_1) (\tau -x_a x_2 )^2} .
\label{eqn:cross-5}
\end{align}
\ \vspace{-0.45cm} \  

\noindent
The variable $\tau$ is given the dimuon mass and 
the center-of-mass energy squared as $\tau=M_{\mu\mu}^2/s=Q^2/s$,
$\bm q_T^{\,2}$ is the dimuon transverse momentum squared,
 $\phi$ is its azimuthal angle, $y$ is the rapidity 
in the center-of-mass frame,
and the color factor is $C_F=(N_c^2-1)/(2N_c)$ with $N_c=3$.
The quark and antiquark distribution functions in the proton,
$q_A (x_a)$ and $\bar q_A (x_a)$, are well known, so that
the gluon transversity in the deuteron $\Delta_T g (x_b)$ 
could be extracted from the cross-section measurement.

\vspace{0.00cm}
\begin{table}[t]
\scriptsize
\begin{center}
\renewcommand{\arraystretch}{1.6} 
\bottomcaption{
\small
Longitudinal, transverse, and linear polarizations
of a spin-1 hadron in terms of the polarization vectors $\bm E$
and the spin-vector/tensor parameters
($S_{T}^x$, $S_{T}^y$, $\cdots$)
\cite{g-tran-drell-yan-2020}.}
\label{table:polarizations}
\begin{supertabular}{|l|c|ccccc|} \hline
Polarizations & $\bm E$ &  $S_T^x$
     \s0 $S_T^y$ \s0 $S_L$  \s0 $S_{LL}$ \s0 $S_{TT}^{xx}$ \\ \hline
Longitudinal $+z$  & $\frac{1}{\sqrt{2}} (-1,\, -i,\, 0)$  &
       0    \s0   0     \s0  $+$1 \s0 $+\frac{1}{2}$ \s0  0  \\ \hline
Longitudinal $-z$ & $\frac{1}{\sqrt{2}} (+1,\, -i,\, 0)$ &
       0    \s0   0     \s0  $-$1 \s0 $+\frac{1}{2}$ \s0   0  \\ \hline
Transverse $+x$ & $\frac{1}{\sqrt{2}} (0,\, -1,\, -i)$ &
         $+$1    \s0   0     \s0  0  \s0 $-\frac{1}{4}$  \s0  $+\frac{1}{2}$ \\ \hline
Transverse $-x$ & $\frac{1}{\sqrt{2}} (0,\, +1,\, -i)$ &
         $-1$    \s0   0     \s0  0  \s0 $-\frac{1}{4}$ \s0 $+\frac{1}{2}$ \\ \hline
Transverse $+y$ & $\frac{1}{\sqrt{2}} (-i,\, 0,\, -1)$ &
         0   \s0   $+$1      \s0  0  \s0 $-\frac{1}{4}$ \s0 $-\frac{1}{2}$ \\ \hline
Transverse $-y$ & $\frac{1}{\sqrt{2}} (-i,\, 0,\, +1)$ &
         0   \s0   $-1$     \s0  0  \s0 $-\frac{1}{4}$ \s0 $-\frac{1}{2}$ \\ \hline
Linear  $x$  &  $(1,\, 0,\, 0)$ &
         0   \s0   0      \s0  0  \s0 $+\frac{1}{2}$ \s0 $-1$ \\ \hline
Linear  $y$  &  $(0,\, 1,\, 0)$ &
         0   \s0   0     \s0  0  \s0 $+\frac{1}{2}$ \s0 $+1$ \\ \hline
\end{supertabular}
\end{center}
\vspace{-0.70cm}
\normalsize
\end{table}

\begin{figure}[t]
 \vspace{-0.00cm}
  \begin{center}
   \includegraphics[width=6.0cm]{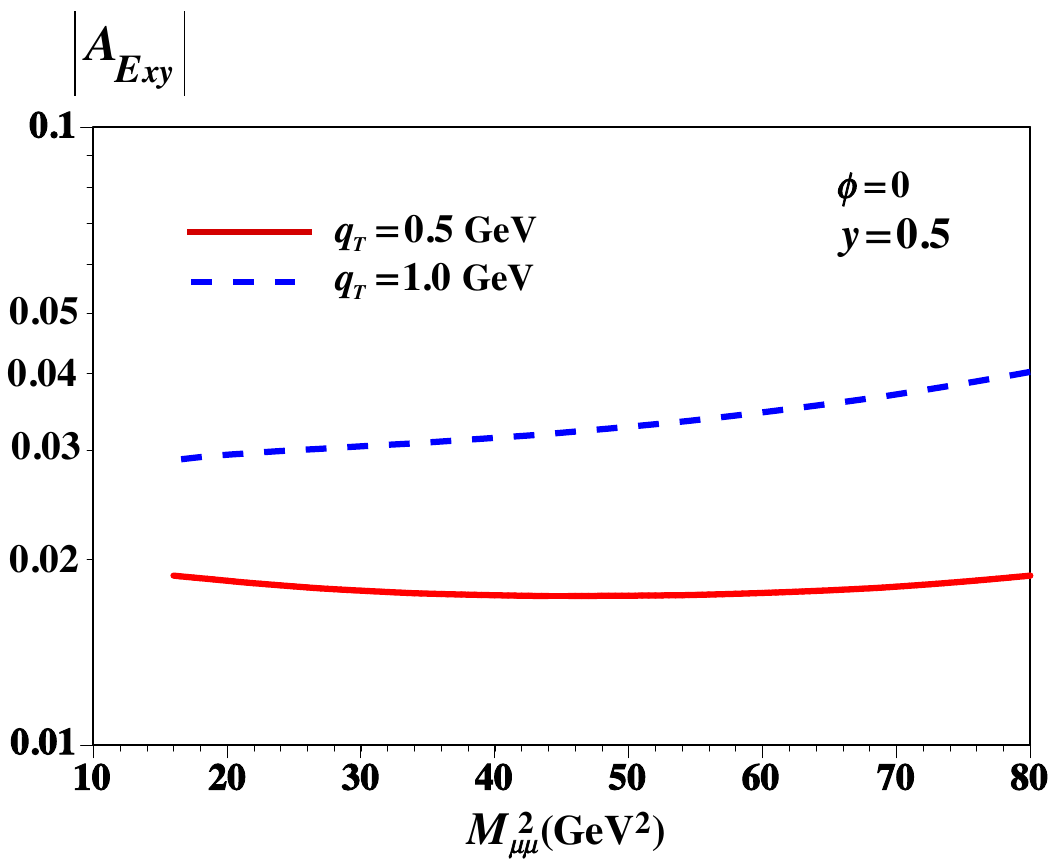}
  \end{center}
\vspace{-0.3cm}
\caption{Polarization asymmetry $ | A_{E_{xy}} |$ in the proton-deuteron
Drell-Yan cross section.}
\label{fig:A-Exy}
\vspace{-0.30cm}
\end{figure}

The cross sections are calculated by using the CTEQ 14 
PDFs for the unpolarized PDFs of the proton,
and the gluon transversity in the deuteron is assumed
as the longitudinally-polarized gluon distribution 
given by the NNPDF1.1 because there is no available 
$\Delta_T g (x_b)$. It is possibly an overestimation 
of the cross section.
The polarization asymmetry 
$A_{E_{xy}}\equiv d\sigma (E_x-E_y)/d\sigma (E_x+E_y)$
is shown for $p_p = 120$ GeV, $\phi=0$, $y=0.5$, 
and $q_T=0.5$ or 1.0 GeV as the function $M_{\mu\mu}^2$
in Fig.\,\ref{fig:A-Exy} by considering the Fermilab experiment.
The asymmetry is of the order of a few percent.
This kind of experiment is also possible at NICA
\cite{NICA-2021}.
Possibly, the JLab, Fermilab, NICA, and EIC experiments could
create a new field of hadron physics 
if a finite gluon transversity is found.

\vspace{-0.30cm}
\section{TMDs and PDFs up to twist 4}
\label{tmds}
\vspace{-0.15cm}

\subsection{Classifications of TMDs and PDFs}
\label{tmds-pdfs-4}
\vspace{-0.15cm}

\begin{table*}[t!]
\vspace{-0.00cm}
\begin{minipage}[c]{0.333\textwidth}
   \includegraphics[width=5.3cm]{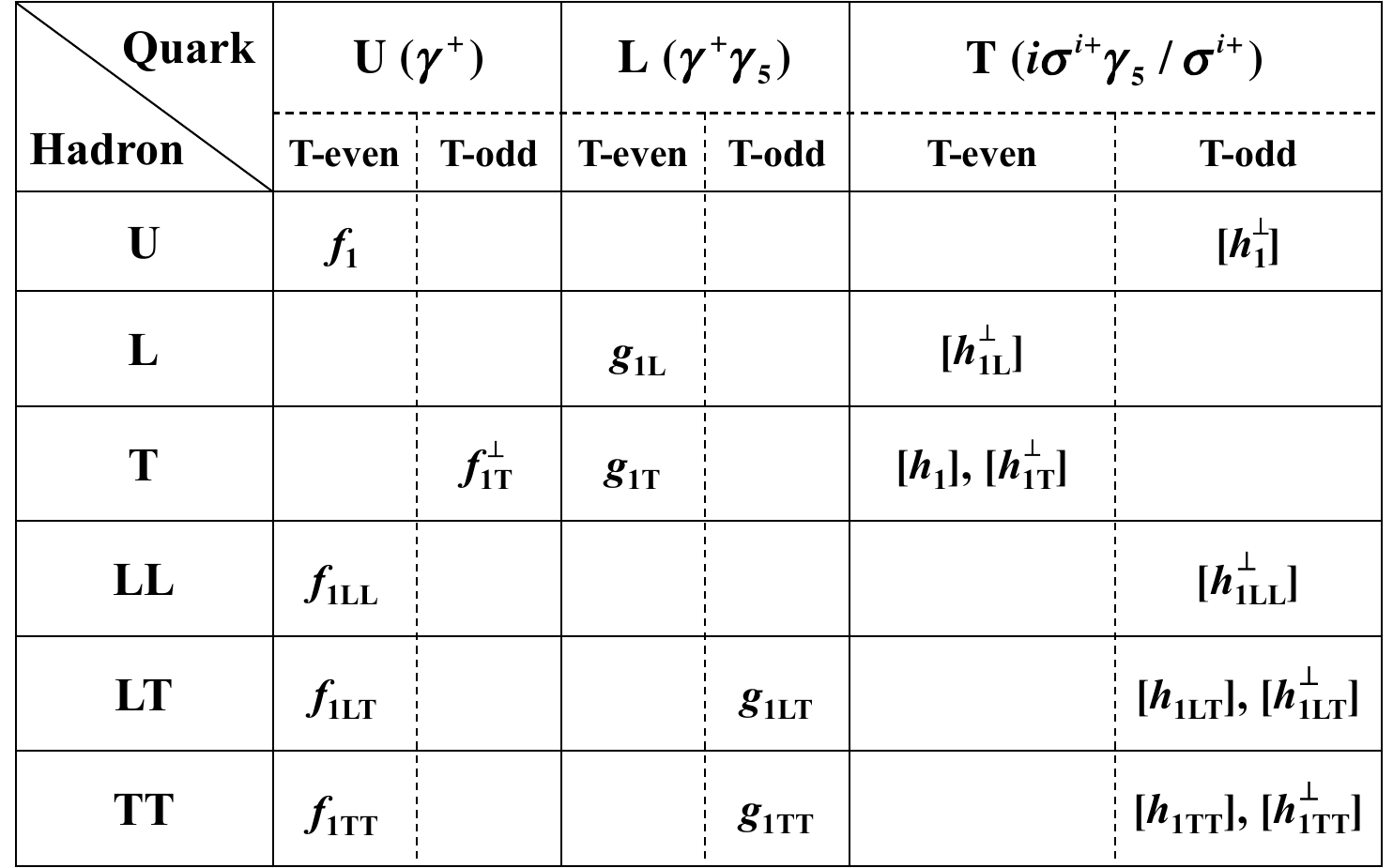}
      \caption{Twist-2 TMDs.}
      \label{tab:TW2-TMDs}
\end{minipage} 
\hspace{-0.20cm}
\begin{minipage}[c]{0.333\textwidth}
\vspace{0.075cm}
   \includegraphics[width=5.65cm]{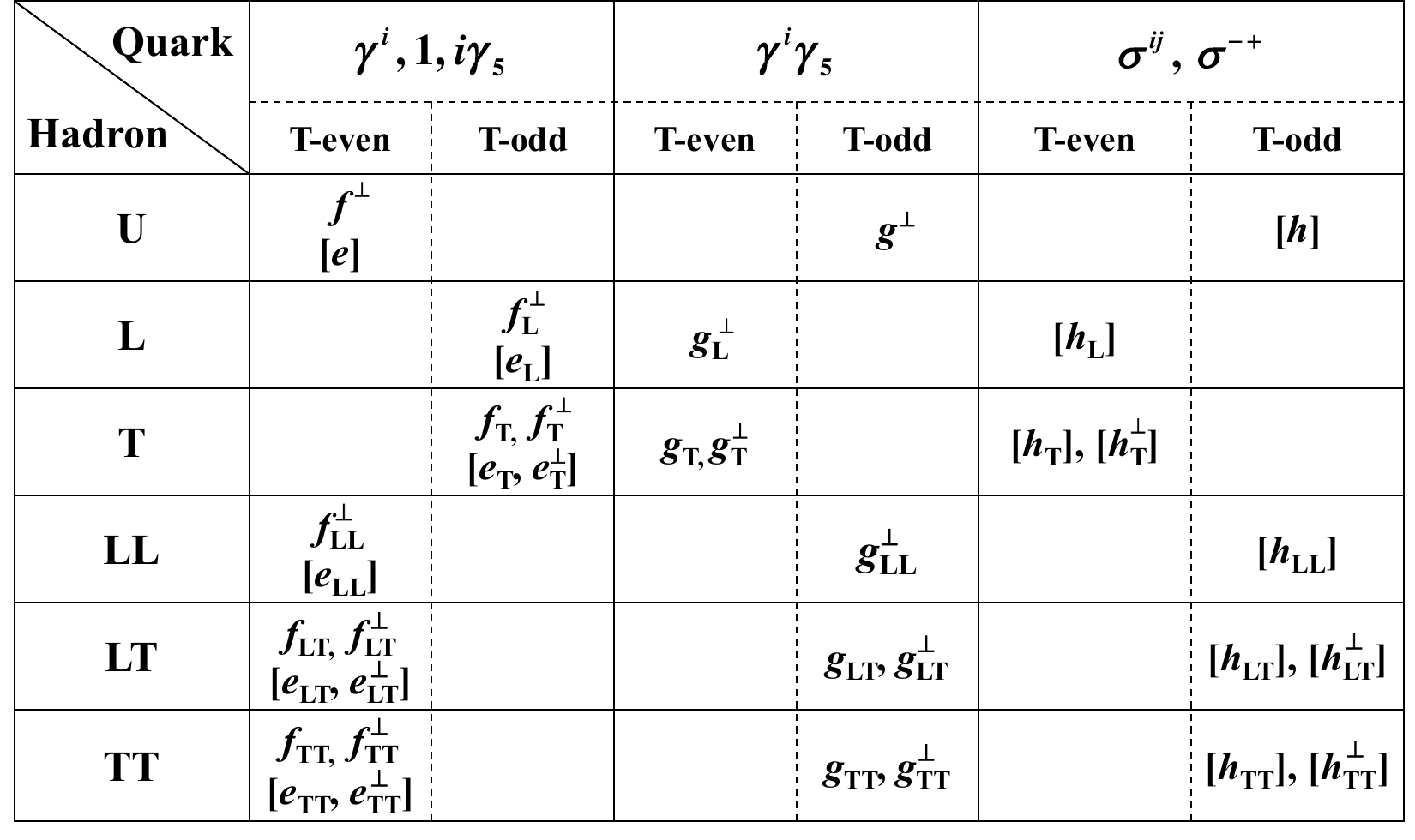}
      \caption{Twist-3 TMDs.}
      \label{tab:TW3-TMDs}
\end{minipage} 
\begin{minipage}[c]{0.333\textwidth}
\vspace{0.050cm}
\hspace{0.08cm}
   \includegraphics[width=5.62cm]{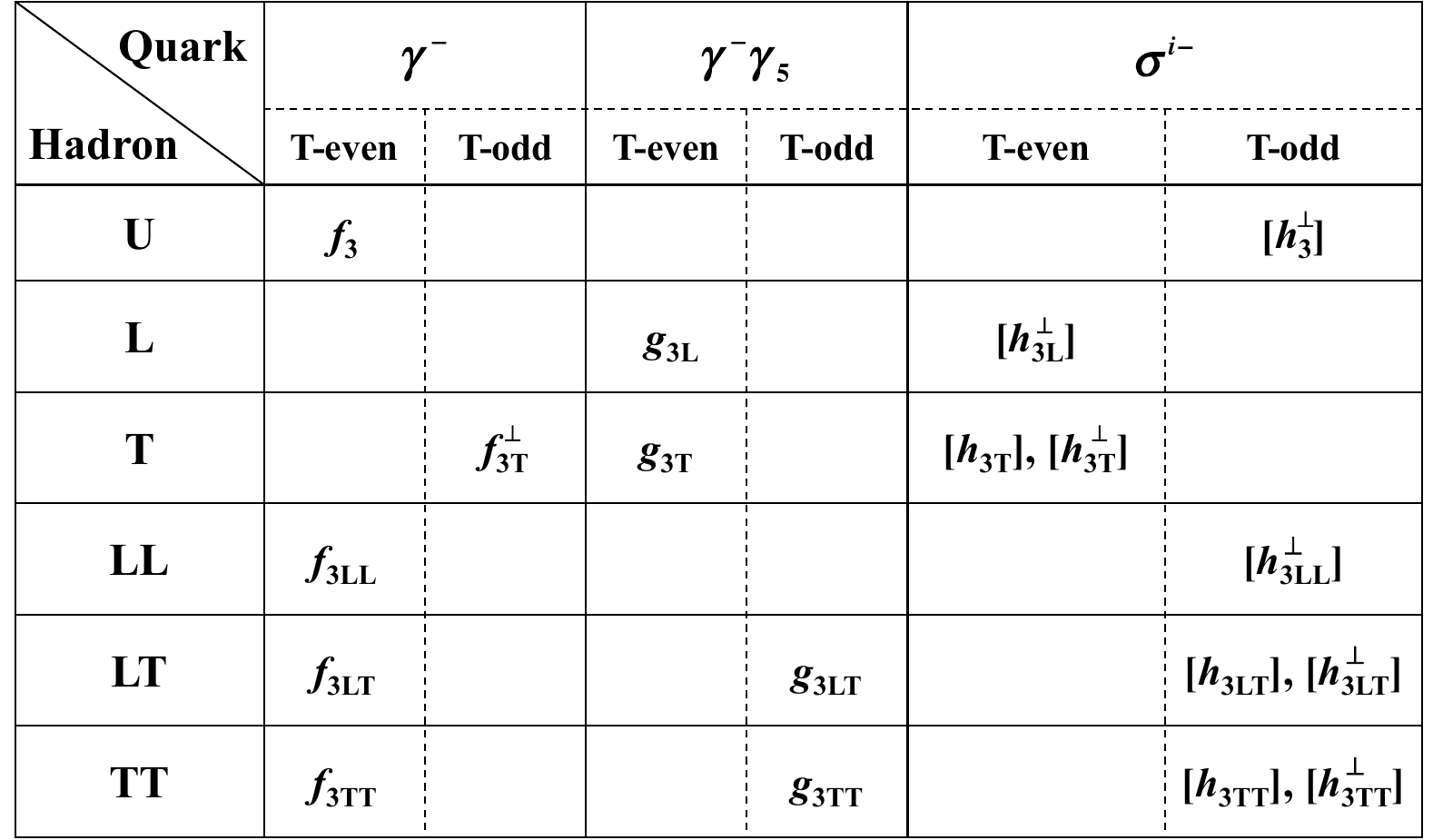}
      \caption{Twist-4 TMDs.}
      \label{tab:TW4-TMDs}
\end{minipage} 
\vspace{-0.00cm}
\end{table*}

\begin{table*}[t!]
\vspace{-0.30cm}
\begin{minipage}[c]{0.333\textwidth}
   \includegraphics[width=5.6cm]{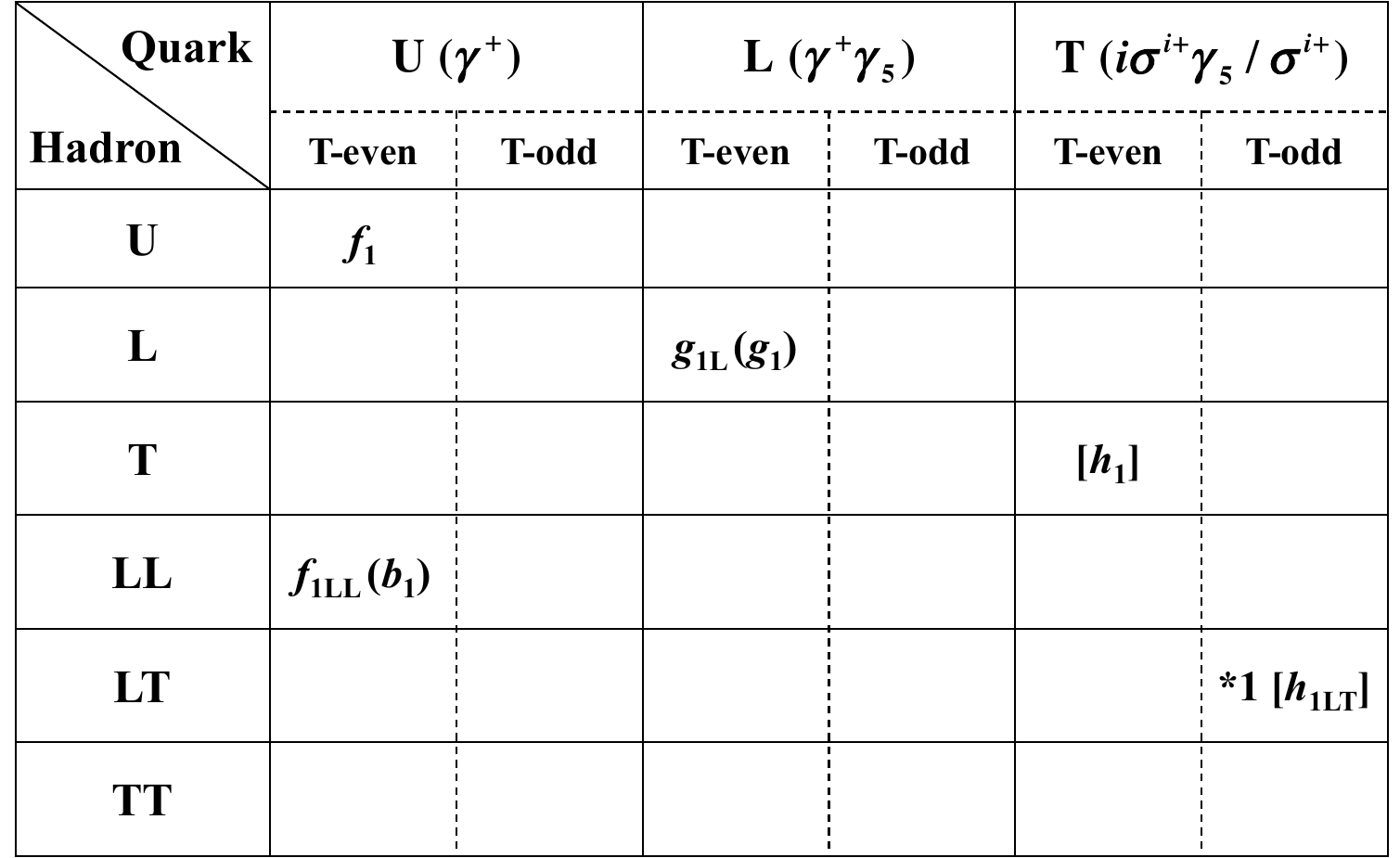}
      \caption{Twist-2 PDFs.}
      \label{tab:TW2-PDFs}
\end{minipage} 
\hspace{0.00cm}
\begin{minipage}[c]{0.333\textwidth}
\vspace{0.000cm}
   \includegraphics[width=5.55cm]{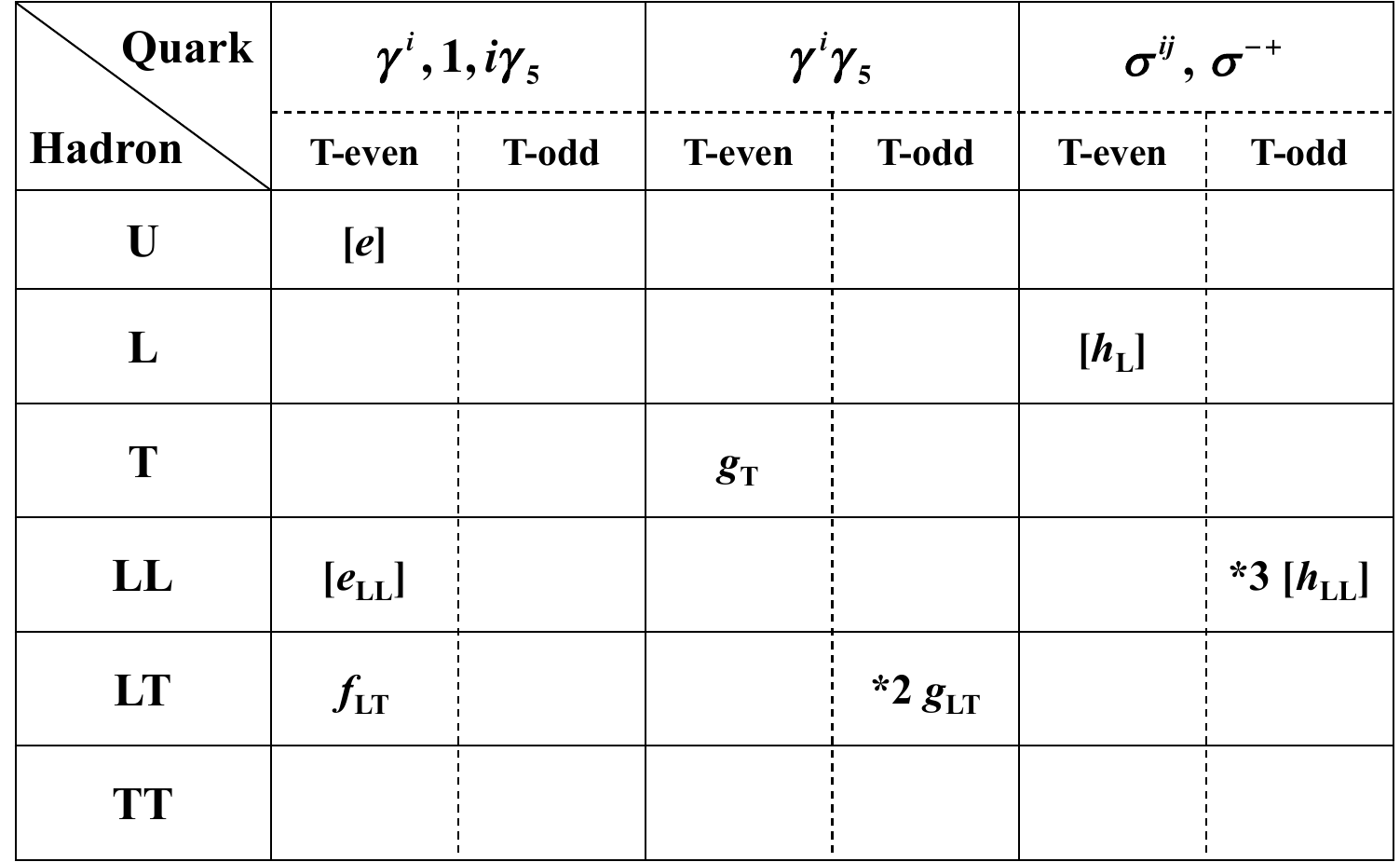}
      \caption{Twist-3 PDFs.}
      \label{tab:TW3-PDFs}
\end{minipage} 
\hspace{-0.25cm}
\begin{minipage}[c]{0.333\textwidth}
\vspace{-0.700cm}
\hspace{0.08cm}
   \includegraphics[width=5.60cm]{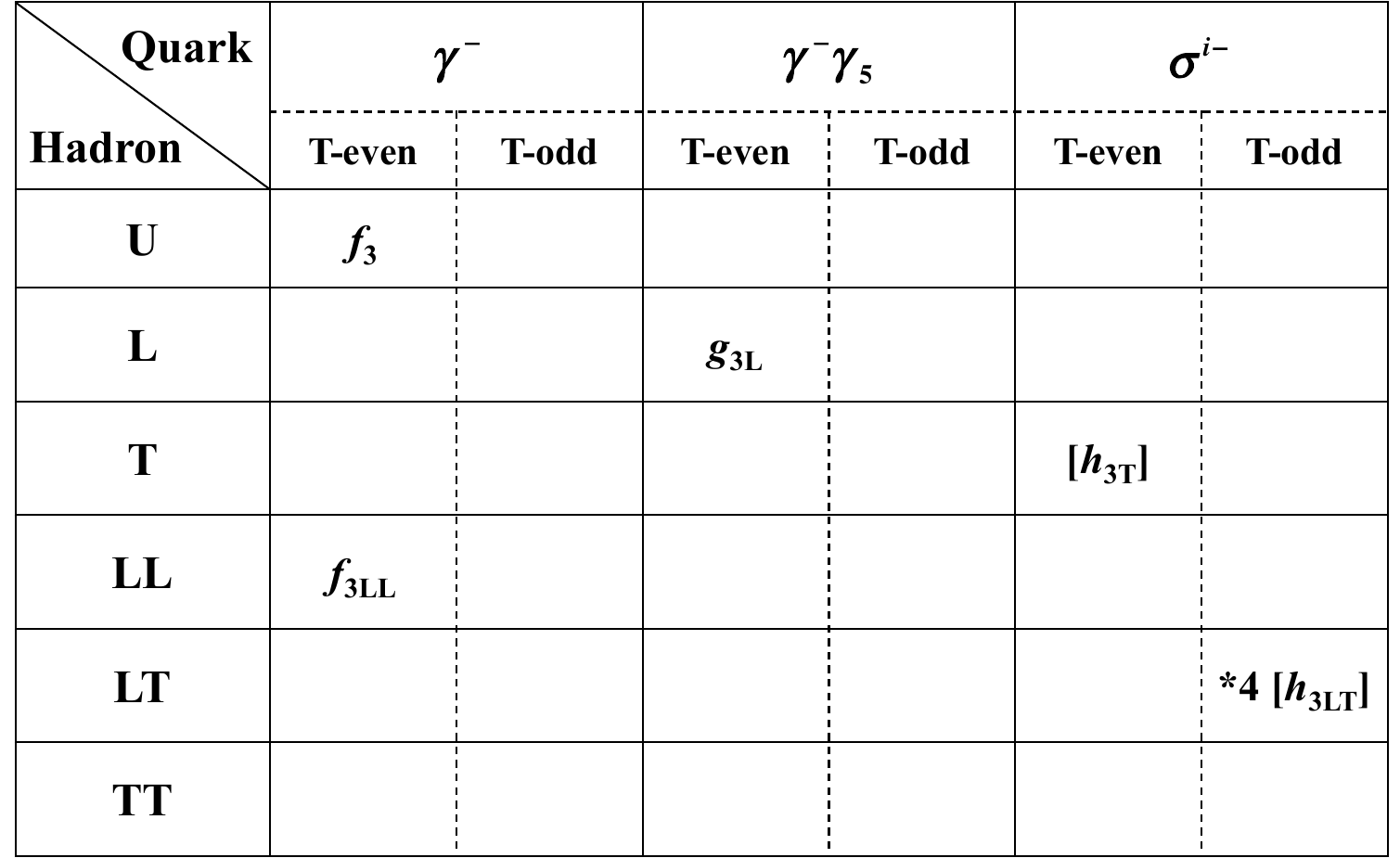}
      \caption{Twist-4 PDFs.}
      \label{tab:TW4-PDFs}
\vspace{-0.70cm}
\end{minipage} 
\vspace{-0.30cm}
\end{table*}

Possible TMDs and PDFs up to twist 4 were listed in 
Ref.\,\cite{spin-1-TMDs-2021} for spin-1 hadrons.
They were derived in general from a correlation function
defined by the amplitude to extract a parton from a hadron and 
then to insert it into the hadron at a different space-time point as
\vspace{-0.10cm}
\begin{align}
& \Phi_{ij}^{[c]} (k, P, T ) 
= \int  \! \frac{d^4 \xi}{(2\pi)^4} \, e^{ i k \cdot \xi}
\nonumber \\[-0.10cm]
& \hspace{1.7cm}
\times
\langle \, P , T \left | \, 
\bar\psi _j (0) \,  W^{[c]} (0, \xi)  
 \psi _i (\xi)  \, \right | P, \,  T \, \rangle ,
\label{eqn:correlation-q}
\\[-0.70cm] \nonumber
\end{align} 
where $P$ and $T$ indicate the spin-1 hadron momentum
and tensor polarization. 
The vector polarization is not explicitly written.
The $k$ and $\psi$ are quark momentum and field, and
$ W^{[c]} (0, \xi)$ is the gauge link with the integral path $c$.
The covariant form of the tensor polarization of
Eq.\,(\ref{eqn:spin-1-vector-tensor})
is written as
\vspace{-0.10cm}
\begin{align}
&
T^{\mu\nu}  = \frac{1}{2} \left [ \frac{4}{3} S_{LL} \frac{(P^+)^2}{M^2} 
               \bar n^\mu \bar n^\nu 
          - \frac{2}{3} S_{LL} ( \bar n^{\{ \mu} n^{\nu \}} -g_T^{\mu\nu} )
\right.
\nonumber \\[-0.10cm]
& \! 
\left.
\! + \! 
\frac{1}{3} S_{LL} \frac{M^2}{(P^+)^2}n^\mu n^\nu
\! + \! 
\frac{P^+}{M} \bar n^{\{ \mu} S_{LT}^{\nu \}}
\! - \!
\frac{M}{2 P^+} n^{\{ \mu} S_{LT}^{\nu \}}
\! + \! 
S_{TT}^{\mu\nu} \right ] ,
\label{eqn:spin-1-tensor-1}
\\[-0.80cm] \nonumber
\end{align}
where $n$ and $\bar n$ are the lightcone vectors 
$n^\mu =(1,0,0,-1)$ $/\sqrt{2}$ and $\bar n^\mu =(1,0,0,1)/\sqrt{2}$,
$g_T^{\mu\nu}$ is given by $g_T^{11}=g_T^{22}=-1$ 
and other components vanishing,
$a^{\{ \mu} b^{\nu \}}$ indicates 
$a^{\{ \mu} b^{\nu \}} = a^\mu b^\nu + a^\nu b^\mu$,
and $M$ is the hadron mass.

For projecting out possible TMDs and PDFs, the correlation function is
decomposed by considering the Hermiticity and parity invariance as
\cite{spin-1-TMDs-2021}
\vspace{-0.00cm}
\begin{align}
\Phi ^{[c]} (k, P, T ) & = \frac{A_{13}}{M}  T_{kk} 
+ \cdots
+ \frac{A_{20}}{M^2} \varepsilon^{\mu\nu P k}  \gamma_{\mu} \gamma_5 T_{\nu k}
\nonumber \\[-0.10cm]
& \ 
+ \frac{B_{21}M}{P\cdot n} T_{kn}  
+ \cdots
+ \frac{B_{52}M}{P\cdot n } \sigma_{\mu k}  T^{\mu n} ,
\label{eqn:cork4}
\\[-0.60cm] \nonumber
\end{align} 
where the contraction is defined as
$X_{\mu k} \equiv X_{\mu \nu} k^{\nu}$.
The expression in Eq.\,(\ref{eqn:cork4}) extends
the twist-2 expression 
(8 terms of $A_{13}, \cdots , A_{20}$)
in Ref.\,\cite{BM-2000}
with the additional 32 $n$-dependent terms ($B_{21}, \cdots , B_{52}$)
for studying twist-3 and twist-4 TMDs and PDFs.
Equation (\ref{eqn:cork4}) indicates that 
the correlation function depends on the lightcone vector $n$
due to the integral path of the gauge link.
Then, the TMDs and collinear correlation functions are given by
integrating it over the quark momenta as
\vspace{-0.20cm}
\begin{align}
\Phi^{[c]} (x, k_T, P, T ) &  = \int dk^+ dk^- \, 
               \Phi^{[c]} (k, P, T) \, \delta (k^+ -x P^+) ,
\nonumber \\
\Phi (x, P, T ) & 
  =  \int d^2 k_T \, \Phi^{[c]} (x, k_T, P, T ) .
\label{eqn:correlation-pdf}
\\[-0.60cm] \nonumber
\end{align}

Possible TMDs and PDFs are obtained by the traces of
these correlation functions with $\gamma$ matrices ($\Gamma$) as
$ \Phi^{\left[ \Gamma \right]} \equiv 
\frac{1}{2} \, \text{Tr} \left[ \, \Phi \Gamma \, \right] $.
The twist-2 TMDs and PDFs are defined by the traces
$\Phi^{ [ \gamma^+ ] }$,
$\Phi^{ [ \gamma^+ \gamma_5 ] }$, and
$\Phi^{ [ i \sigma^{i+} \gamma_5 ] }$
(or $\Phi^{ [ \sigma^{i+} ] }$),
the twist-3 functions are by
$\Phi^{ [ \gamma^i ] }$,
$\Phi^{\left[{\bf 1}\right]}$,
$\Phi^{\left[i\gamma_5\right]}$
$\Phi^{ [\gamma^{i}\gamma_5 ]}$
$\Phi^{ [ \sigma^{ij} ]}$,
and $\Phi^{ [ \sigma^{-+} ] }$, 
where $i$ and $j$ are transverse indices ($i,j=1$ or 2),
and the twist-4 functions are by
$\Phi^{[\gamma^-]}$,
$\Phi^{[\gamma^- \gamma_5]}$, and $\Phi^{[\sigma^{i-}]}$
as shown in Tables 
\ref{tab:TW2-TMDs}--\ref{tab:TW4-PDFs}.
As an example for the twist-3 TMDs, we have the relation
\begin{align}
& 
\Phi^{ [ \gamma^i ] } (x, k_T, T)
\! = \! 
\frac{M}{P^+} \bigg [  f^{\perp}_{LL}(x, k_T^{\, 2})  S_{LL} \frac{k_T^i}{M}
\! + \! f^{\,\prime} _{LT} (x, k_T^{\, 2})S_{LT}^i 
\nonumber \\[-0.10cm]
& \ \hspace{1.0cm}
- f_{LT}^{\perp}(x, k_T^{\, 2}) \frac{ k_{T}^i  S_{LT}\cdot k_{T}}{M^2} 
- f_{TT}^{\,\prime} (x, k_T^{\, 2}) \frac{S_{TT}^{ i j} k_{T \, j} }{M} 
\nonumber \\[-0.10cm]
& \ \hspace{1.0cm}
+ f_{TT}^{\perp}(x, k_T^{\, 2}) \frac{k_T\cdot S_{TT}\cdot k_T}{M^2} 
       \frac{k_T^i}{M} \bigg ] .
\label{eqn:cork-3-1a}
\end{align} 
In Table \ref{tab:TW3-TMDs}, the TMDs without $^\prime$ are shown
by defining 
$
F (x, k_T^{\, 2}) \equiv F^{\,\prime} (x, k_T^{\, 2})
 - (k_T^{\, 2} /(2M^2)) \, F^{\perp} (x, k^{\, 2}_T) 
$
where $k_T^{\, 2}= - \bm k_T^{\, 2}$.
From these studies, we found that
there are 40 TMDs in total, and they are 
10, 20, and 10 tensor-polarized TMDs 
at twists 2, 3, and 4, respectively, as
\begin{align}
& \text{Twist-2 TMD:}\ f_{1LL},\, f_{1LT},\, f_{1TT},\, g_{1LT},\, g_{1TT},
\nonumber \\[-0.10cm]
& \ \hspace{2.24cm}
      h_{1LL}^\perp,\, h_{1LT},\, h_{1LT}^\perp,\, h_{1TT},\, h_{1TT}^\perp ,
\nonumber \\
& \text{Twist-3 TMD:}\ f_{LL}^\perp,\, e_{LL},\,  
      f_{LT},\, f_{LT}^\perp,\, e_{1T},\, e_{1T}^\perp,\,
      f_{TT},\, f_{TT}^\perp,
\nonumber \\[-0.10cm]
& \ \hspace{2.24cm}      
      e_{TT},\ e_{TT}^\perp,\,
      g_{LL}^\perp,\, g_{LT},\, g_{LT}^\perp,\, g_{TT},\, g_{TT}^\perp,
\nonumber \\[-0.10cm]
& \ \hspace{2.24cm}
      h_{1L},\, h_{LT},\, h_{LT}^\perp,\, h_{TT},\, h_{TT}^\perp,
\nonumber \\
& \text{Twist-4 TMD:}\, f_{3LL},\, f_{3LT},\, f_{3TT},\, g_{3LT},\, g_{3TT},
\nonumber \\[-0.10cm]
& \ \hspace{2.24cm}
      h_{3LL}^\perp,\, h_{3LT},\, h_{3LT}^\perp,\, h_{3TT},\, h_{3TT}^\perp .
\label{eqn:spin-1-tmds-2-3-4}
\end{align} 
In Tables \ref{tab:TW2-TMDs}, \ref{tab:TW3-TMDs}, and \ref{tab:TW4-TMDs},
the TMDs are classified by chiral even/odd and time-reversal even/odd.
Chiral-odd distributions are shown with the square brackets $[\ ]$,
and the distributions without the bracket are chiral-even ones. 
The polarizations U, L, and T indicate
unpolarized, longitudinally polarized,
and transversely polarized, respectively, and
the tensor polarizations are shown by LL, LT, and TT.
The asterisks $*1$, $*2$, $*3$, $*4$ mean that the PDFs
$h_{1LT} (x)$, $g_{LT} (x)$, $h_{LL} (x)$, $h_{3LT} (x)$
vanish because of the time-reversal invariance;
however, the corresponding fragmentation functions
$H_{1LT} (z)$, \\
\noindent
$G_{LT} (z)$, $H_{LL} (z)$, $H_{3LT} (z)$
should exist as collinear fragmentation functions
\cite{spin-1-TMDs-2021,Ji-FFs-1994}, and
finite transverse momentum moments could exist
even for the T-odd TMDs.
The time-reversal invariance should be satisfied for the collinear PDFs,
so that there are sum rules for the T-odd TMDs
by the integral over the transverse momentum $\vec k_T$ as
\begin{align}
& \int \! d^2 k_T \, h_{1LT} (x, k_T^{\, 2}) 
= \int \! d^2 k_T \, g_{LT} (x, k_T^{\, 2}) 
\nonumber \\[-0.20cm]
& \ 
= \int \! d^2 k_T \, h_{LL} (x, k_T^{\, 2}) 
= \int \! d^2 k_T \, h_{3LT}(x, k_T^{\, 2})  = 0 .
\label{eqn:TMD-sum}
\end{align} 
The possible PDFs were obtained by integrating the TMDs 
over $\bm k_T$ up to twist 4 as
\vspace{-0.00cm}
\begin{align}
& \ \ \ \text{Twist-2 PDF:}\ f_{1LL}, 
\nonumber \\[-0.10cm]
& \ \ \ \text{Twist-3:}\ \hspace{0.87cm} e_{LL},\ f_{LT}, 
\nonumber \\[-0.10cm]
& \ \ \ \text{Twist-4:}\ \hspace{0.85cm} f_{3LL}. 
\label{eqn:pdfs-2-3-4}
\end{align} 
The distribution $f_{1LL}$ corresponds to 
the tensor-polarized distribution $\delta_T q$ 
in Sec.\,\ref{spin-1-sfs}
by the relation $f_{1LL} = - (2/3)$ $\delta_T q$.
There are a twist-2 sum rule and useful relations
among these PDFs and multiparton distribution functions
\cite{twist-2-relations-2021,eq-motion-2022}
as discussed in Sec.\,\ref{useful-relations}.
In this way, we are ready to investigate the structure functions
of spin-1 hadrons up to twist 4 in the similar way as for
those of the spin-1/2 nucleon.

\vspace{-0.30cm}
\subsection{Relations of twist 2 and from equation of motion}
\label{useful-relations}
\vspace{-0.15cm}

\begin{table*}[hbt]
\vspace{-0.00cm}
\begin{minipage}[c]{0.333\textwidth}
   \includegraphics[width=5.3cm]{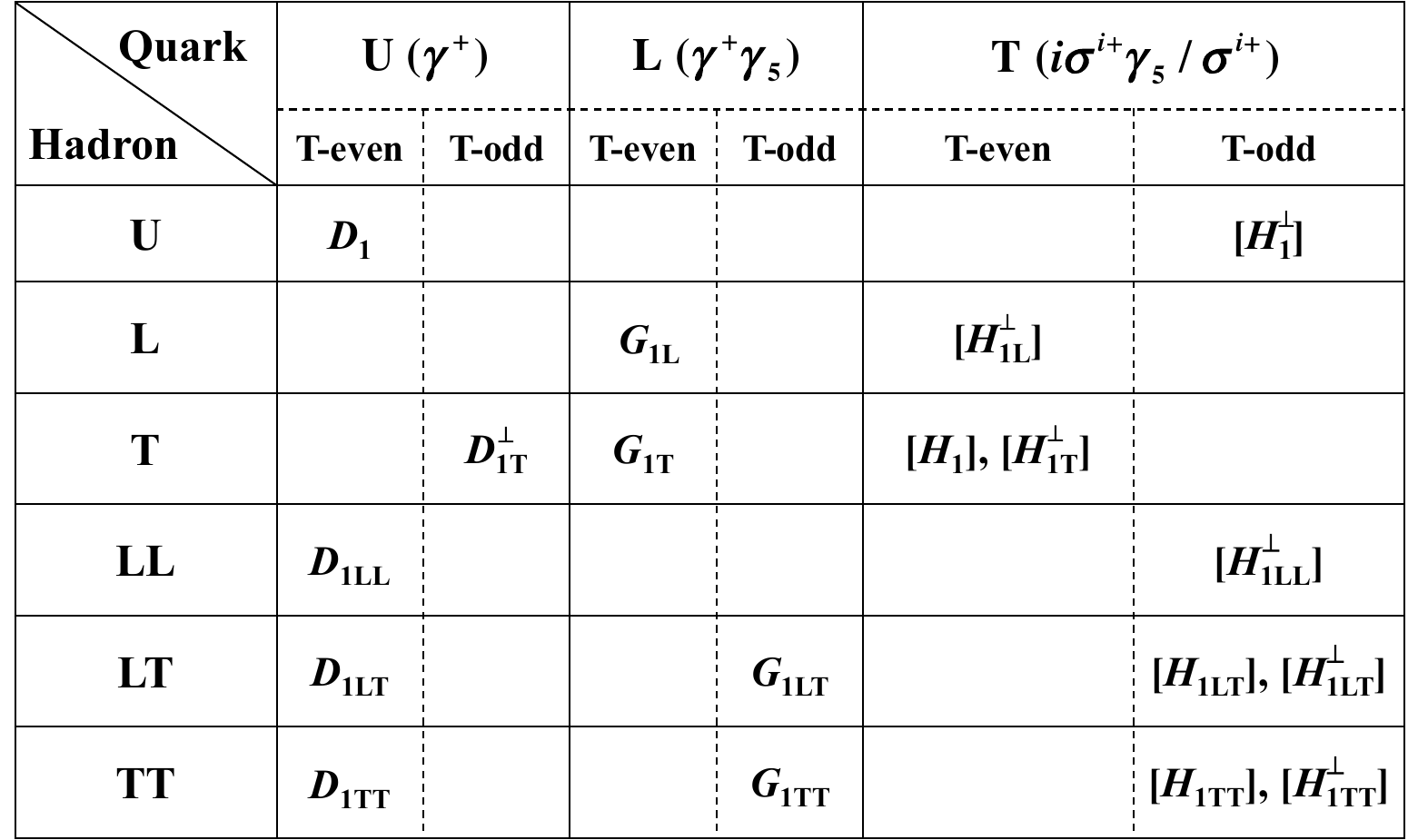}
      \caption{Twist-2 TMD FFs.}
      \label{tab:TW2-TMD-FFs}
\end{minipage} 
\hspace{-0.38cm}
\begin{minipage}[c]{0.333\textwidth}
\vspace{0.075cm}
   \includegraphics[width=5.8cm]{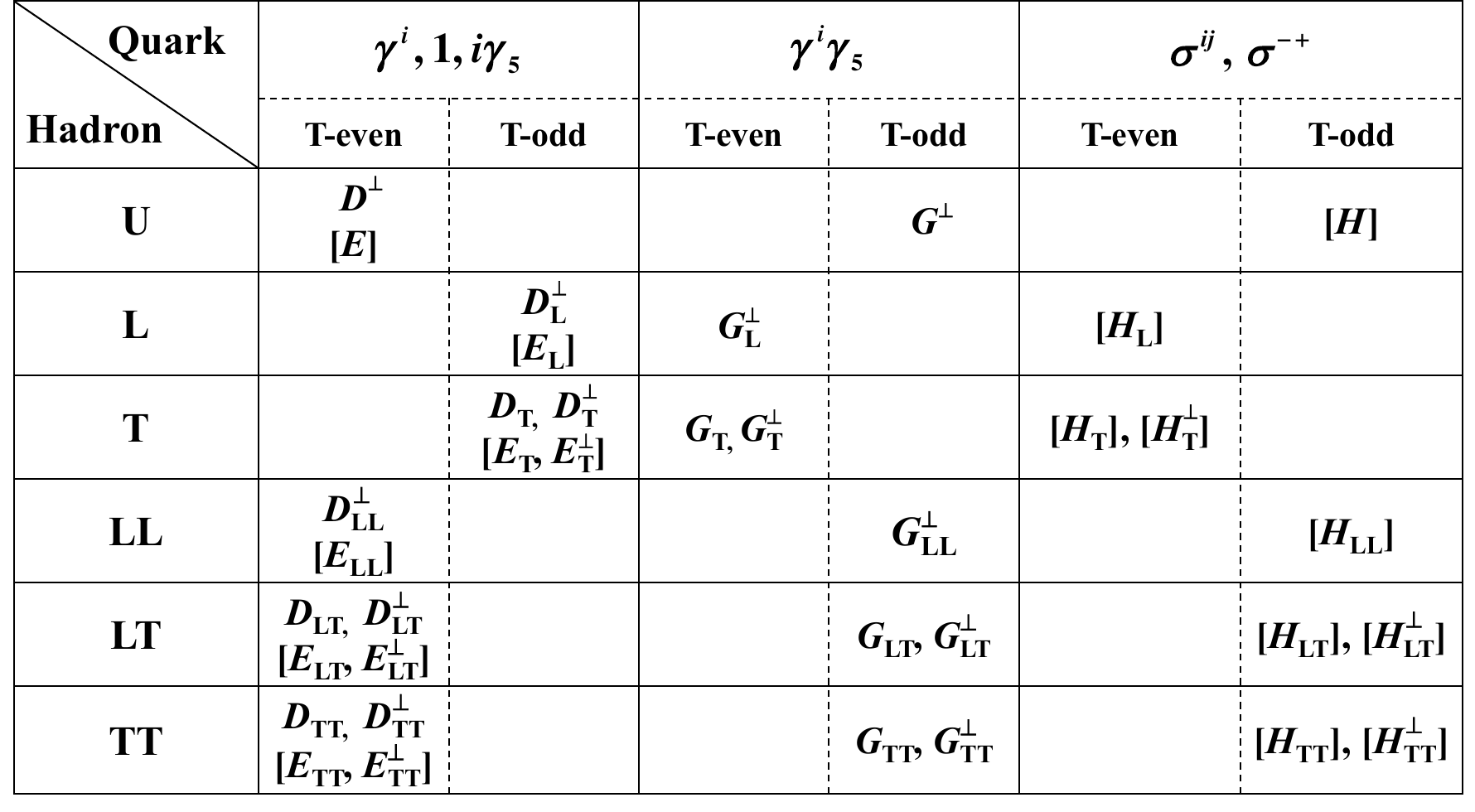}
      \caption{Twist-3 TMD FFs.}
      \label{tab:TW3-TMD-FFs}
\end{minipage} 
\hspace{-0.05cm}
\begin{minipage}[c]{0.333\textwidth}
\vspace{0.050cm}
\hspace{0.08cm}
   \includegraphics[width=5.7cm]{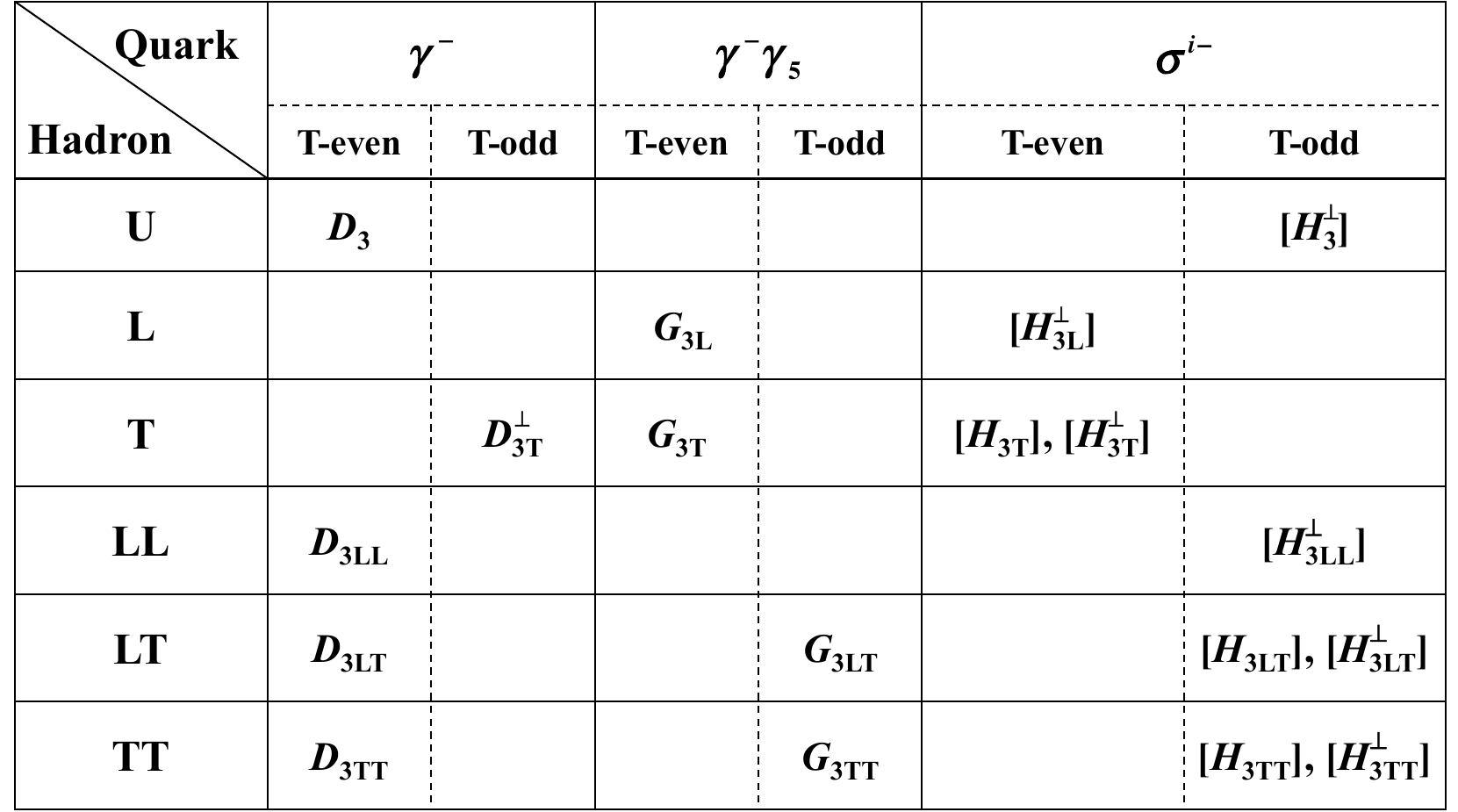}
      \caption{Twist-4 TMD FFs.}
      \label{tab:TW4-TMD-FFs}
\end{minipage} 
\vspace{-0.00cm}
\end{table*}

\begin{table*}[hbt]
\vspace{-0.00cm}
\begin{minipage}[c]{0.333\textwidth}
   \includegraphics[width=5.55cm]{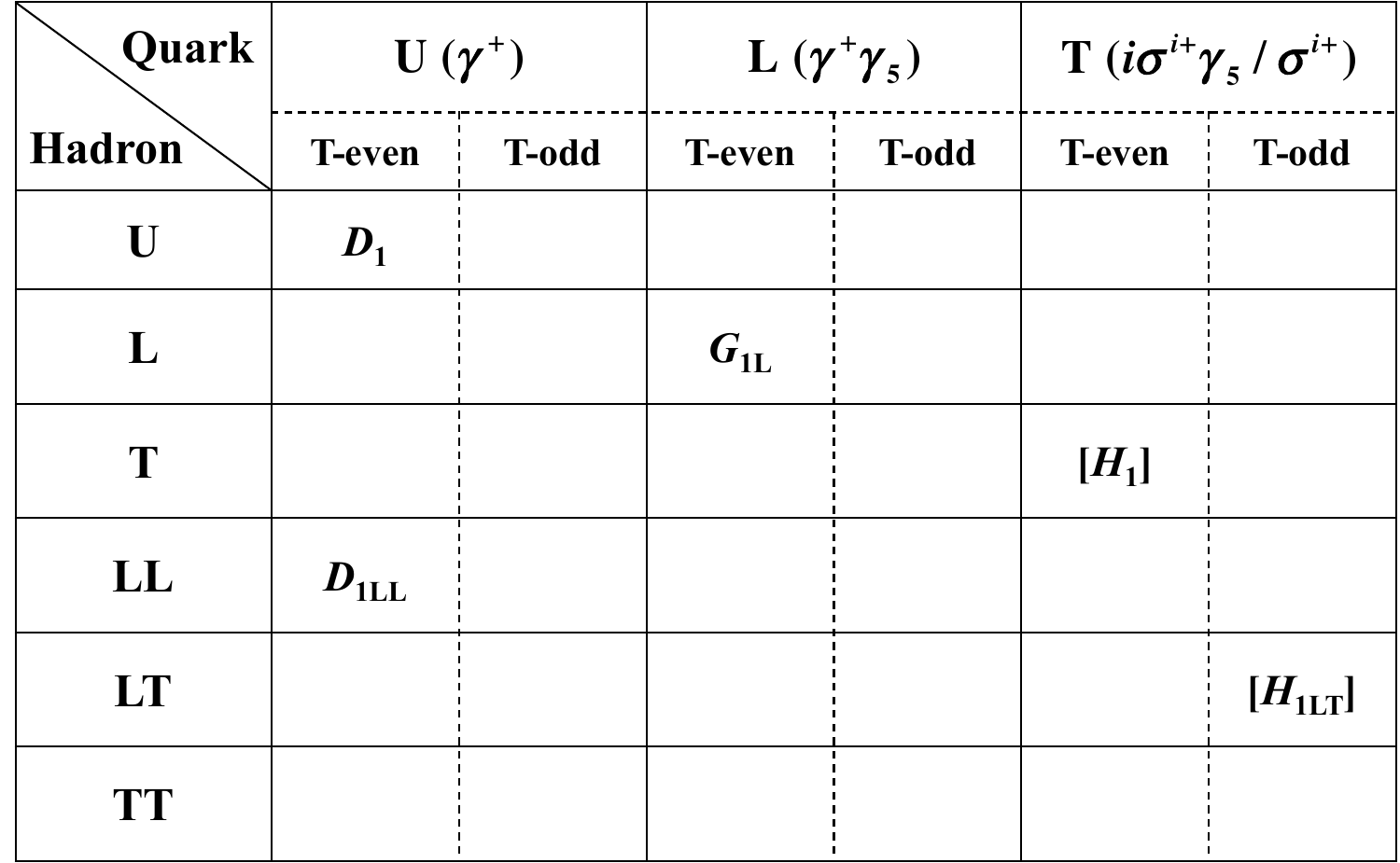}
      \caption{Twist-2 FFs.}
      \label{tab:TW2-FFs}
\end{minipage} 
\hspace{0.00cm}
\begin{minipage}[c]{0.333\textwidth}
\vspace{0.010cm}
   \includegraphics[width=5.55cm]{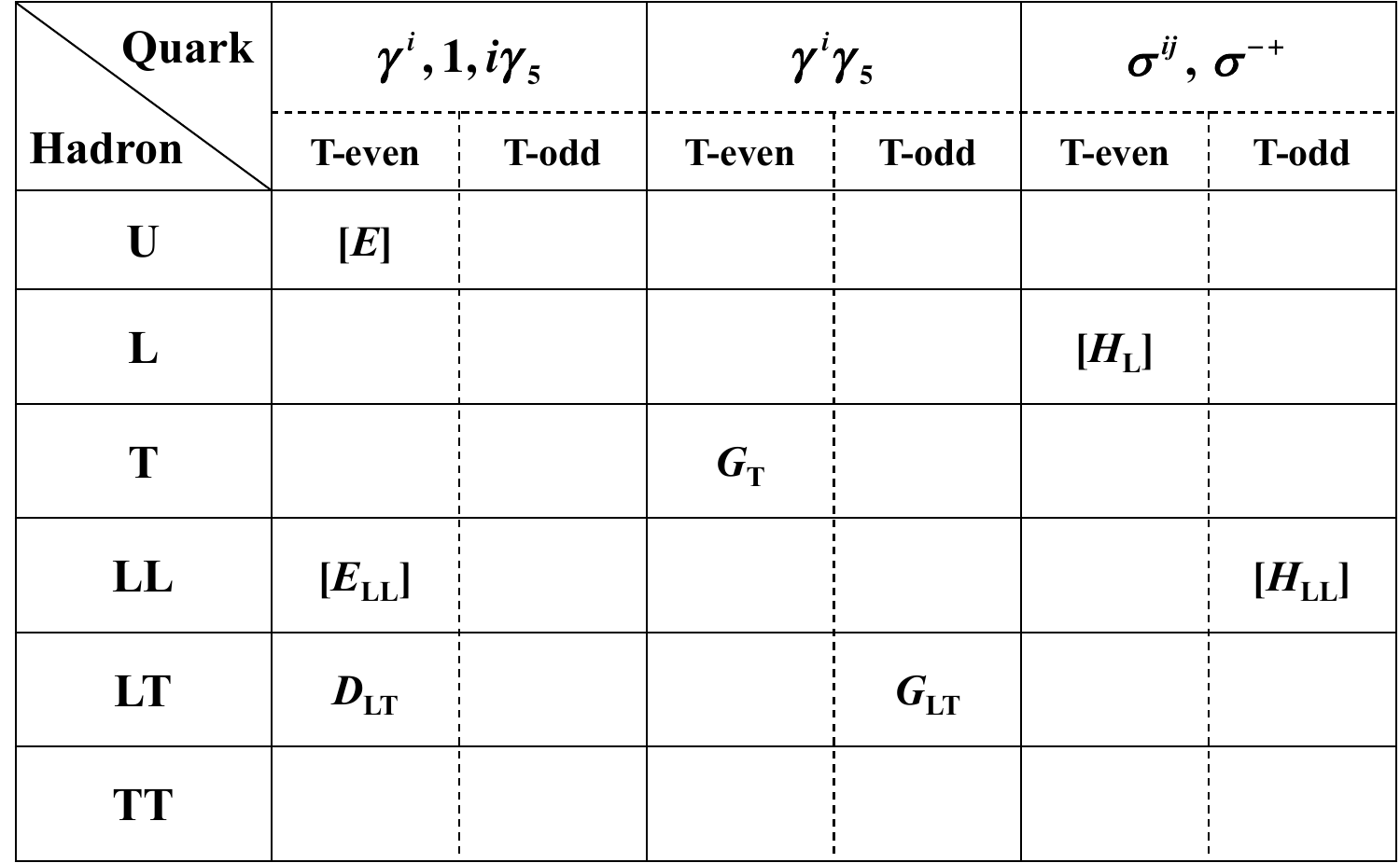}
      \caption{Twist-3 FFs.}
      \label{tab:TW3-FFs}
\end{minipage} 
\hspace{-0.20cm}
\begin{minipage}[c]{0.333\textwidth}
\vspace{0.050cm}
\hspace{0.08cm}
   \includegraphics[width=5.55cm]{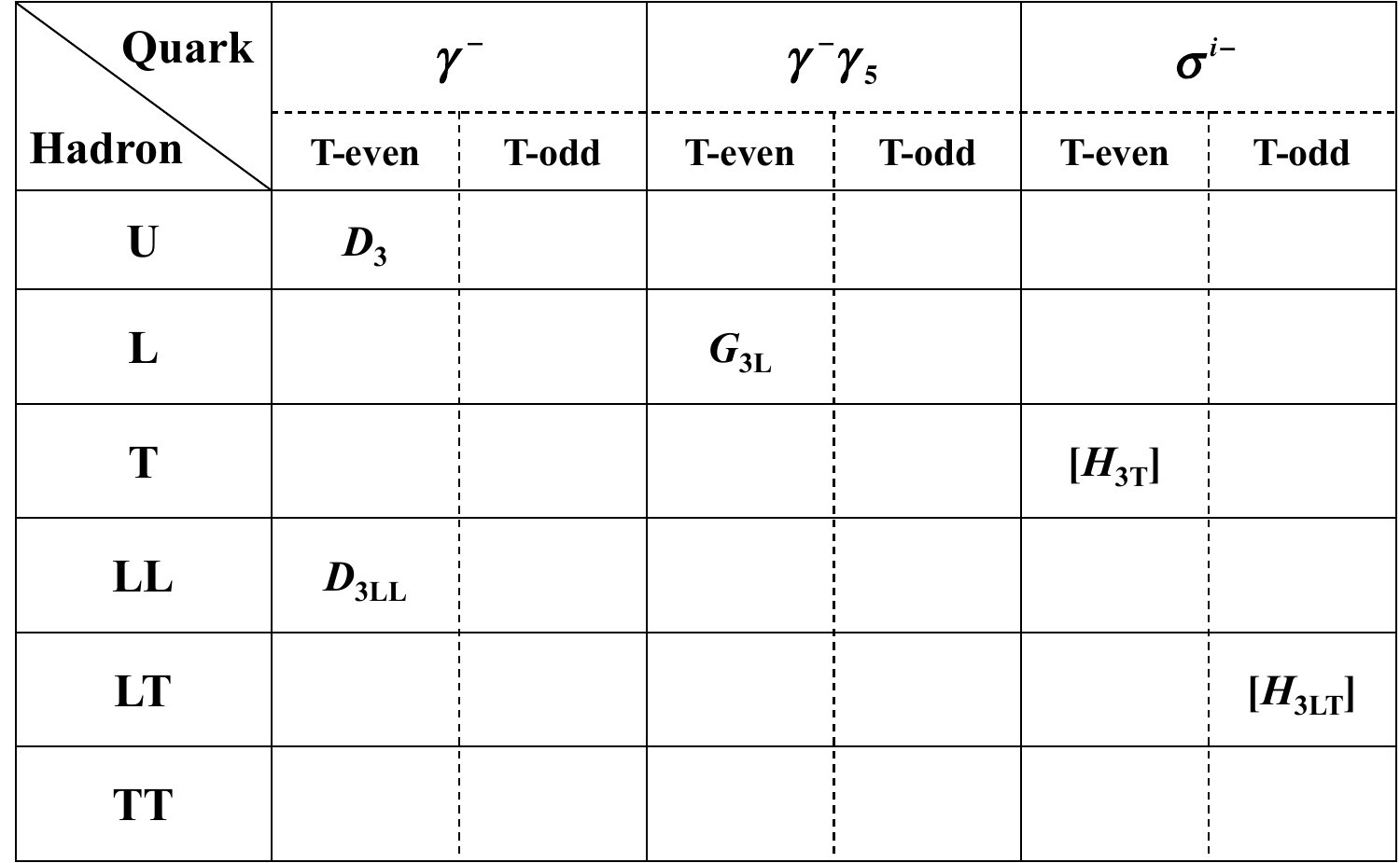}
      \caption{Twist-4 FFs.}
      \label{tab:TW4-FFs}
\end{minipage} 
\vspace{-0.10cm}
\end{table*}

Because the higher-twist PDFs were obtained for 
the tensor-polarized spin-1 hadron as explained 
in Sec.\,\ref{tmds-pdfs-4}, it is interesting
to investigate whether there are useful relations among them
to find their magnitudes and functional forms.
In the spin-1/2 nucleon, there are a twist-2 relation 
called the Wandzura-Wilczek (WW) relation 
and the Burkhardt-Cottingham (BC) sum rule.
Such relations were investigated 
for the twist-2 and twist-3 tensor-polarized PDFs 
$f_{1LL}$ and $f_{LT}$ \cite{twist-2-relations-2021}
of spin-1 hadrons.
We obtained a similar equation to the WW relaton by using
the operator product expansion and defining
multiparton distribution functions for twist-3 terms as
\begin{align}
\! 
f_{LT}(x)= \frac{3}{2} \int^{\epsilon (x)}_x \! \frac{dy}{y} f_{1LL}(y)
           + \int^{\epsilon (x)}_x \! \frac{dy}{y} f_{LT}^{(HT)}(y) ,
\label{eqn:flt}
\end{align}
where $\epsilon (x)=1$ ($-1$) at $x>0$ ($x<0$),
and the last term is the twist-3 term (HT: higher twist)
given by the multiparton distribution functions.
The PDFs $f^+ (x)$ are defined by $f^+ (x) = f(x) + \bar f(x)$ 
in the range $0 \le x \le 1$. 
Then, the $f_{1LL}^+$ is related to $b_1$ by
$b_1^{q+\bar q} = - (3/2) f_{1LL}^+$.
If the higher-twist term is neglected, Eq.\,(\ref{eqn:flt})
becomes
\begin{align}
f_{LT}^+(x)= \frac{3}{2} \int^1_x 
\frac{dy}{y} \, f_{1LL}^+(y) ,
\end{align}
which indicates the twist-2 part of $f_{LT}$ is expressed 
by an integral of the twist-2 distribution $f_{1LL}$ (or $b_1$).
This equation was written in the similar form with
the WW relation by defining $f_{2LL}$ as
$f_{2LL} = \frac{2}{3} f_{LT}-f_{1LL}$:
\begin{align}
f_{2LT}^+ (x)=-f_{1LL}^+ (x)+ \int^1_x \frac{dy}{y} f_{1LL}^+ (y) .
\nonumber 
\end{align}
By integrating this equation over $x$, the BC-like sum rule
was obtained 
\begin{align}
& \int_0^1 dx \, f_{2LT}^+(x) =0 .
\nonumber 
\end{align}
If the $b_1$ sum rule of Eq.\,(\ref{eqn:b1-sum-parton-4})
is used for $f_{1LL}$ ($b_1$), namely
$\int dx f_{1LL}^+ (x) = 0$  ($\int dx b_1^{q+\bar q} (x) = 0$),
by assuming vanishing tensor-polarized antiquark distributions, 
the sum rule 
\begin{align}
\int_0^1 dx \, f_{LT}^+(x) =0 ,
\end{align}
is satisfied for $f_{LT}$ itself.
In these studies, we found the existence of
the tensor-polarized multiparton distribution functions
$F_{LT} (x,y)$, $G_{LT} (x,y)$,
$H_{LL}^\perp (x,y)$, and $H_{TT} (x,y)$.

Other useful relations were derived by using the equation of motion 
for quarks \cite{eq-motion-2022}. 
The transverse momentum moments of the TMDs are defined by 
\begin{align}
f^{\, (1)} (x) = \int \! d^2 k_T \frac{\vec k_T^{\,2}}{2 M^2} \, f(x,k_T^2).
\end{align}
The first relation is for the twist-3 PDF $f_{LT}$,
the trasverse-momentum moment PDF $f_{1LT}^{\,(1)}$, and 
the multiparton distribution functions $F_{G,LT}$ and $G_{G,LT}$:
\begin{align}
& x f_{LT}(x) - f_{1LT}^{\,(1)}(x)
\nonumber \\
&  \ \ \ \ \ \ \ 
- {\cal P}  \int_{-1}^1  dy \,
\frac{F_{G,LT}(x, y) + G_{G,LT} (x, y) }{x-y} = 0 ,
\end{align}
where ${\cal P}$ is the principle integral.
Next, the second relation is 
for the twist-3 PDF $e_{LL}$, the twist-2 PDF $f_{1LL}$,
and the multiparton distribution function $H_{G,LL}^\perp$:
\begin{align}
x \, e_{LL}(x) - 2 {\cal P} \int_{-1}^1 dy \, 
 \frac{H_{G,LL}^\perp (x, y)}{x-y} 
-\frac{m}{M} f_{1LL} (x) 
=0 ,
\end{align}
\ \vspace{-0.45cm} \  

\noindent
where $m$ is the quark mass.
The third relation is the Lorentz-invariance relation:
\begin{align}
\frac{d f_{1LT}^{\,(1)}(x) }{dx} - f_{LT}(x) & + \frac{3}{2} f_{1LL}(x)
\nonumber \\
& 
- 2 {\cal P} \int_{-1}^1 dy \, \frac{F_{G,LT} (x, y)}{(x-y)^2} =0 .
\end{align}
Relations among the multiparton distribution functions
$F_{D/G,LT} (x, y)$, $G_{D/G,LT} (x, y)$,
$H_{D/G,LL}^\perp (x, y)$, 
and \\ 
\noindent
$H_{D/G,TT} (x, y)$ were also obtained.

\vspace{-0.30cm}
\section{Fragmentation functions}
\label{fragmentation}
\vspace{-0.15cm}

The collinear fragmentation functions (FFs) for spin-1 hadrons
were investigated up to twist 4 in Ref.\,\cite{Ji-FFs-1994}.
However, consistent derivations on the TMD FFs of the spin-1 hadrons 
were restricted to the twist-2 until recently,
and the TMDs FFs were obtained up to twist 4 
in Ref.\,\cite{spin-1-TMDs-2021}. 
The collinear FFs and TMD FFs are obtained from the PDFs and TMDs, 
respectively, simply by changing the variables and function names as
\cite{BM-2000,spin-1-TMDs-2021}
\vspace{-0.40cm}
\begin{align}
& \ \hspace{-0.00cm}
\text{Kinematical variables:}   \ \  
x, k_T, S, T, M, n, \gamma^+, \sigma^{i+}
\nonumber \\[-0.15cm]
& \ \hspace{3.0cm}
\Rightarrow \ 
 z, k_T, S_h, T_h, M_h, \bar n, \gamma^-, \sigma^{i-},
\nonumber \\
& \ \hspace{-0.00cm}
\text{Distribution functions:}  \ \ f, g, h, e \hspace{2.30cm}
\nonumber \\[-0.15cm]
& \ \hspace{-0.10cm}
\Rightarrow 
\text{Fragmentation functions:} \ 
D, G, H, E .
\label{eqn:tmd-fragmentation}
\end{align} 
For spin-1 hadrons, the twist-2, 3, and 4 TMD FFs are listed in 
Tables \ref{tab:TW2-TMD-FFs}, \ref{tab:TW3-TMD-FFs}, and \ref{tab:TW4-TMD-FFs},
and the twist-2, 3, and 4 collinear FFs are listed in 
Tables \ref{tab:TW2-FFs}, \ref{tab:TW3-FFs}, and \ref{tab:TW4-FFs}.
As mentioned in Sec.\,\ref{tmds} for the asterisks of
Tables \ref{tab:TW2-PDFs}, \ref{tab:TW3-PDFs}, and \ref{tab:TW4-PDFs},
the corresponding T-odd collinear fragmentation functions
exist, although the T-odd PDFs do not exist, because
the time-reversal invariance does not have to be imposed in the FFs.
These collinear and TMD FFs could be investigated experimentally
by future experiments.

Similarly to the PDFs, there are relations among
the FFs and the multiparton distributions \cite{Song-FFs-2023}. 
First, the equation of motion for quarks leads to 
the relations
\begin{align}
& E_{LL}(z)+ i H_{LL}(z)-\frac{m_q}{M} z D_{1LL}(z)
\nonumber \\[-0.00cm]
& \,
= 2 z \left[  -i H^{\perp (1)}_{1LL}(z)+ \mathcal{P}\int^{\infty}_{z} 
\frac{dz_1}{(z_1)^2} \frac{H_{G,LL}^\perp(z,z_1)}
{\frac{1}{z}-\frac{1}{z_1}} \right],
\end{align}
\vspace{-0.30cm}
\begin{align}
& D_{LT}(z) + i G_{LT}(z) +\frac{im_q}{M} z H_{1LT}(z) 
\nonumber \\[-0.00cm]
& \ \ \ 
= -z 
\left [
i G_{1LT}^{(1)}(z) - \int^{\infty}_{z} \frac{dz_1}{(z_1)^2} 
\frac{G_{G,LT}(z,z_1)}{\frac{1}{z}-\frac{1}{z_1}} \right ]
\nonumber \\[-0.00cm]
& \ \ \ \ \ \,
- z \left[ D_{1LT}^{(1)}(z) +\int^{\infty}_{z} \frac{dz_1}{(z_1)^2}
 \frac{D_{G,LT}(z,z_1)}{\frac{1}{z}-\frac{1}{z_1}} \right ],
\end{align}
\vspace{-0.30cm}
\begin{align}
i H^{(1)}_{1TT}(z)+\int^{\infty}_{z} \frac{dz_1}{(z_1)^2} 
\frac{H_{G,TT}(z,z_1)}{\frac{1}{z}-\frac{1}{z_1}} = 0 .
\end{align}
There are also Lorentz-invariance relations
\begin{align}
& \frac{3}{2} D_{1LL}(z) -D_{LT}(z)  
- z \left ( 1-z \frac{d}{dz} \right ) D_{1LT}^{(1)}(z)
\nonumber \\[-0.00cm]
& \ \ \ 
=  - 2 \int_z^{\infty} \frac{dz_1}{(z_1)^2} 
 \frac{\mathrm{Re} \left [ D_{G,LT}(z,z_1) \right] }
 { (\frac{1}{z}-\frac{1}{z_1})^2},
\end{align}
\vspace{-0.30cm}
\begin{align}
& H_{LL}(z) +2 H_{1LT}(z)
+ z \left ( 1-z \frac{d}{dz} \right ) H_{1LL}^{\perp (1)}(z)
\nonumber \\[-0.00cm]
& \ \ \ 
=  - 2 \int_z^{\infty} \frac{dz_1}{(z_1)^2} 
\frac{\mathrm{Im} \left [ H_{G,LL}^\perp (z,z_1) \right] }
{ (\frac{1}{z}-\frac{1}{z_1})^2},
\end{align}
\vspace{-0.30cm}
\begin{align}
& G_{LT}(z) + z \left ( 1-z \frac{d}{dz} \right ) G_{1LT}^{(1)}(z)
\nonumber \\[-0.00cm]
& \ \ \ 
=  - 2 \int_z^{\infty} \frac{dz_1}{(z_1)^2} 
\frac{\mathrm{Im} \left [ G_{G,LT}(z,z_1) \right] }
{ (\frac{1}{z}-\frac{1}{z_1})^2}   .
\end{align}
These relations are useful in studying the twist-3 FFs in future.

\vspace{0.30cm}
Note added:
In this paper, we did not have space to discuss
the generalized parton distributions for spin-1 hadrons.
We list references \cite{BCDP-gpd-2001,CP-trans-gpd-2018}
for understanding the current status.
There were lightcone-model studies on the spin-1 $\rho$-meson GPDs
\cite{BD-rho-GPDs,Kumar-rho-GPDs-2019}
and TMDs \cite{Chao-vector-m-TMDs-2022,Kaur-rho-TMDs-2024}.

\vspace{-0.30cm}
\section{Summary}
\label{summary}
\vspace{-0.15cm}

Polarized structure functions have not been investigated 
for the spin-1 hadrons except for the HERMES measurement on $b_1$; 
however, the situation will change in a few years due to 
the JLab project on the tensor-polarized spin-1 deuteron.
Furthermore, they could be investigated at Fermilab
by the proton-deuteron Drell-Yan processes, and 
the polarized deuteron beams will be available at NICA and EICs.
Theoretical tools are ready up to twist 4 for the collinear PDFs,
the TMDs, the collinear fragmentation functions, and 
TMD fragmentation functions.
For example, the tensor-polarized structure function $b_1$ and 
the gluon transversity associated with the deuteron's
linear polarization could indicate a new hadron physics
beyond the basic deuteron description in terms of 
a bound system of a proton and a neutron.
From the future experimental measurements, there is a possibility
that a new field of hadron physics could be created.

\vspace{-0.40cm}
\section{Acknowledgements}
\vspace{-0.15cm}

SK was partially supported by 
Japan Society for the Promotion of Science (JSPS) Grants-in-Aid
for Scientific Research (KAKENHI) Grant Number 24K07026.

\vspace{-0.40cm}
\bibliographystyle{epj}
\vspace{-0.00cm}

\bibliography{spin-1-pdfs-tmds-ffs}

\begin{thebibliography}{49}

\bibitem{eic-2022}
R.~{Abdul Khalek}, {\it et al}., Nucl. Phys. A \textbf{1026}, 122447 (2022)

\bibitem{EicC-2021}
D.~Anderle, {\it et al}., Front. Phys. \textbf{16}, 64701 (2021)

\bibitem{hermes-b1-2005}
A.~Airapetian, {\it et al}. (HERMES Collaboration), Phys. Rev. Lett.
  \textbf{95}, 242001 (2005)

\bibitem{Jlab-b1}
K.~Allada, {\it et al}., Proposal to Jefferson Lab PAC-51,\\ PR12-13-011,
  https://www.jlab.org/exp\_prog\\ /proposals/13/PR12-13-011.pdf  (2023)

\bibitem{jlab-gluon-trans}
J.~Maxwell, {\it et al}., arXiv: 1803.11206  (2018)

\bibitem{Fermilab-spin}
M.~Brooks, {\it et al}., https://twist.phys.virginia.edu
  /work/E1039proposal\_final.pdf; SpinQuest Collaboration,
  https://spinquest.fnal.gov/  (2017)

\bibitem{Keller-2022}
D.~Keller, arXiv: 2205.01249  (2022)

\bibitem{NICA-2021}
A.~Arbuzov, {\it et al}., Prog. Part. Nucl. Phys. \textbf{119}, 103858 (2021)

\bibitem{fs83}
L.~Frankfurt, M.~Strikman, Nucl. Phys. A \textbf{405}, 557 (1983)

\bibitem{hjm89}
P.~Hoodbhoy, R.~Jaffe, A.~Manohar, Nucl. Phys. B \textbf{312}, 571 (1989)

\bibitem{b1-convolution-2017}
W.~Cosyn, Y.B. Dong, S.~Kumano, M.~Sargsian, Phys. Rev. D \textbf{95}, 074036
  (2017)

\bibitem{JM-g-transversity-1989}
R.~Jaffe, A.~Manohar, Phys. Lett. B \textbf{223}, 218 (1989)

\bibitem{spin-1-projection-2008}
T.Y. Kimura, S.~Kumano, Phys. Rev. D \textbf{78}, 117505 (2008)

\bibitem{tensor-summary-2014}
S.~Kumano, J. Phys. Conf. Ser. \textbf{543}, 012001 (2014)

\bibitem{jm89}
R.~Jaffe, A.~Manohar, Nucl. Phys. B \textbf{321}, 343 (1989)

\bibitem{b1sum}
F.E. Close, S.~Kumano, Phys. Rev. D \textbf{42}, 2377 (1990)

\bibitem{feynman-book}
R.P. Feynman, p.91, Photon-Hadron Interactions, W. A. Benjamin, Inc.  (1972)

\bibitem{KUMANO1998183}
S.~Kumano, Phys. Rep. \textbf{303}, 183 (1998)

\bibitem{GARVEY2001203}
G.~Garvey, J.C. Peng, Prog. Part. Nucl. Phys. \textbf{47}, 203 (2001)

\bibitem{PENG201443}
J.C. Peng, J.W. Qiu, Prog. Part. Nucl. Phys. \textbf{76}, 43 (2014)

\bibitem{ET-1982}
A.V. Efremov, O.V. Teryaev, Sov. J. Nucl. Phys. \textbf{36}, 557 (1982)

\bibitem{miller-2014}
G.A. Miller, Phys. Rev. C \textbf{89}, 045203 (2014)

\bibitem{tensor-pdfs}
S.~Kumano, Phys. Rev. D \textbf{82}, 017501 (2010)

\bibitem{pd-Drell-Yan-tensor-2016}
S.~Kumano, Q.T. Song, Phys. Rev. D \textbf{94}, 054022 (2016)

\bibitem{pd-Drell-Yan}
S.~Hino, S.~Kumano, Phys. Rev. D {\bf 59}, 094026 (1999); {\bf 60}, 054018
  (1999)

\bibitem{pd-Drell-Yan-antiquark}
S.~Kumano, M.~Miyama, Phys. Lett. B \textbf{479}, 149 (2000)

\bibitem{g-tran-drell-yan-2020}
S.~Kumano, Q.T. Song, Phys. Rev. D {\bf 101}, 054011 (2020); {\bf 101}, 094013
  (2020)

\bibitem{Artru-Mekhfi-1990}
X.~Artru, M.~Mekhfi, Z. Phys. C \textbf{45}, 669 (1990)

\bibitem{Mulders-Rodrigues-2001}
P.J. Mulders, J.~Rodrigues, Phys. Rev. D \textbf{63}, 094021 (2001)

\bibitem{Nzar-Hoodbhoy-1992}
M.~Nzar, P.~Hoodbhoy, Phys. Rev. D \textbf{45}, 2264 (1992)

\bibitem{Vogelsang-1998}
W.~Vogelsang, Acta Phys. Pol. B \textbf{29}, 1189 (1998)

\bibitem{Sather-Schmidt-1990}
E.~Sather, C.~Schmidt, Phys. Rev. D \textbf{42}, 1424 (1990)

\bibitem{Detmold-Shanahan-2016}
W.~Detmold, P.E. Shanahan, Phys. Rev. D {\bf 94}, 014507 (2016); Erratum, {\bf
  95}, 079902  (2017)

\bibitem{BM-2000}
A.~Bacchetta, P.J. Mulders, Phys. Rev. D \textbf{62}, 114004 (2000)

\bibitem{Bohr-2016}
D.~Boer, {\it et al}., J. High Energy Phys. \textbf{2016}, 13 (2016)

\bibitem{Meissner-2007}
S.~Mei\ss{}ner, A.~Metz, K.~Goeke, Phys. Rev. D \textbf{76}, 034002 (2007)

\bibitem{QCD-handbook-1995}
G.~Sterman, {\it et al}., Rev. Mod. Phys. \textbf{67}, 157 (1995)

\bibitem{MWZ-2013}
J.P. Ma, C.~Wang, G.P. Zhang, arXiv: 1306.6693  (2013)

\bibitem{spin-1-TMDs-2021}
S.~Kumano, Q.T. Song, Phys. Rev. D \textbf{103}, 014025 (2021)

\bibitem{Ji-FFs-1994}
X.~Ji, Phys. Rev. D \textbf{49}, 114 (1994)

\bibitem{twist-2-relations-2021}
S.~Kumano, Q.T. Song, J. High Energy Phys. \textbf{2021}, 141 (2021)

\bibitem{eq-motion-2022}
S.~Kumano, Q.T. Song, Phys. Lett. B \textbf{826}, 136908 (2022)

\bibitem{Song-FFs-2023}
Q.T. Song, Phys. Rev. D \textbf{108}, 094041 (2023)

\bibitem{BCDP-gpd-2001}
E.R. Berger, F.~Cano, M.~Diehl, B.~Pire, Phys. Rev. Lett. \textbf{87}, 142302
  (2001)

\bibitem{CP-trans-gpd-2018}
W.~Cosyn, B.~Pire, Phys. Rev. D \textbf{98}, 074020 (2018)

\bibitem{BD-rho-GPDs}
B.D. Sun, Y.B. Dong, Phys. Rev. D {\bf 96}, 036019 (2017); {\bf 99}, 016023
  (2019); {\bf 101}, 096008  (2020)

\bibitem{Kumar-rho-GPDs-2019}
N.~Kumar, Phys. Rev. D \textbf{99}, 014039 (2019)

\bibitem{Chao-vector-m-TMDs-2022}
C.~Shi, {\it et al}., Phys. Rev. D \textbf{106}, 014026 (2022)

\bibitem{Kaur-rho-TMDs-2024}
S.~Kaur, {\it et al}., Phys. Lett. B \textbf{851}, 138563 (2024)

\end{thebibliography}

\end{document}